\documentclass[12pt,preprint]{aastex}

\shorttitle{Formation of Self-Gravitating Cores}
\shortauthors{Li et al.}

\begin{document}

\title{The Formation of Self-Gravitating Cores in Turbulent Magnetized Clouds}

%% Use \author, \affil, and the \and command to format
%% author and affiliation information.
%% Note that \email has replaced the old \authoremail command
%% from AASTeX v4.0. You can use \email to mark an email address
%% anywhere in the paper, not just in the front matter.
%% As in the title, you can use \\ to force line breaks.

\author{P. S. Li\altaffilmark{1} and Michael L. Norman\altaffilmark{2}}
\affil{Physics Department, University of California,
    San Diego, CASS/UCSD 0424, 9500 Gilman Drive, La Jolla, CA 92093-0424}

\author{Mordecai-Mark Mac Low\altaffilmark{3}}
\affil{Department of Astrophysics, American Museum of Natural History,
Central Park West at 79th Street, New York, NY 10024-5192}

\and

\author{Fabian Heitsch\altaffilmark{4}}
\affil{Max-Planck-Institut f\"ur Astronomie, K\"onigstuhl 17,
    69117 Heidelberg, Germany}

\altaffiltext{1}{Current address: Astronomy Department, 601 Campbell Hall, University of California at Berkeley, Berkeley, CA 94720-3411. psli@astro.berkeley.edu}
\altaffiltext{2}{mnorman@cosmos.ucsd.edu}
\altaffiltext{3}{mordecai@amnh.org}
\altaffiltext{4}{Current address: Universit\"ats-Sternwarte M\"unchen, Ludwig-Maximilians-Universit\"at M\"unchen, Scheinerstr. 1, D-81679 M\"unchen, Germany.  heitsch@usm.uni-muenchen.de}

\begin{abstract}

We use ZEUS-MP to perform high resolution, three-dimensional, super-Alfv\'{e}nic turbulent simulations in order to investigate the role of magnetic fields in self-gravitating core formation within turbulent molecular clouds.  Statistical properties of our super-Alfv\'{e}nic model without gravity agree with previous similar studies.  Including self-gravity, our models give the following results.  They are consistent with the turbulent fragmentation prediction of the core mass distribution of Padoan \& Nordlund.  They also confirm that local gravitational collapse is not prevented by magnetohydrodynamic waves driven by turbulent flows, even when the turbulent Jeans mass exceeds the mass in the simulation volume.  Comparison of results between 256$^3$ and 512$^3$ zone simulations reveals convergence in the collapse rate.  Analysis of self-gravitating cores formed in the simulation shows that: (1) All cores formed are magnetically supercritical by at least an order of magnitude.  (2) A power law relation between central magnetic field strength and density $B_c \propto \rho_c^{1/2}$ is observed despite the cores being strongly supercritical.  (3) Specific angular momentum $j \propto R^{3/2}$ for cores with radius $R$. (4) most cores are prolate and triaxial in shape, in agreement with the results of Gammie et al.  We find a weak correlation between the minor axis of the core and the local magnetic field in our simulation at late times, different from the uncorrelated results reported by Gammie et al. The core shape analysis and the power law relationship between core mass and radius $M \propto R^{2.75}$ suggest the formation of some highly flattened cores.  We identified twelve cloud cores with disk-like appearance at a later stage of our high-resolution simulation.  Instead of being tidally truncated or disrupted, the core disks survive and flourish while undergoing strong interactions.  We discuss the physical properties of these disk-like cores under the constraints of resolution limits.

\end{abstract}


\section{Introduction}

In the standard theory of isolated star formation, the presence of magnetic fields in molecular clouds plays a vital role.  Based on the Jeans argument, observed high-density molecular clouds should collapse within a few freefall times to form stars.  Instead, the molecular clouds were thought to be quasi-stable.  Turbulence in molecular clouds and magnetic field support were proposed to explain these apparently stable clouds \citep[e.g.][]{mck99,wil00,vaz00}.  Molecular clouds were suggested to be supported magnetostatically \citep[e.g.][]{mou76,fie99} or dynamically by magnetohydrodynamic (MHD) waves, especially Alfv\'{e}n waves \citep[e.g.][]{dew70,shu87}.  Alfv\'{e}n waves are transverse waves and produce perturbation perpendicular to the mean magnetic field.  \citet{mck95} pointed out that Alfv\'{e}n waves can lead to an isotropic turbulence pressure to counteract gravitational collapse, provided that the waves are neither damped nor driven.  However, this mechanism requires a negative radial gradient in wave sources in the cloud in order to support the cloud from collapsing \citep{shu87}.

Observationally, the question of whether magnetic fields are sufficiently strong to support molecular clouds alone from collapsing remains unresolved.  \citet{cru99} concludes that static magnetic fields are insufficient to support the observed clouds and that MHD waves are equally important in cloud energetics.  \citet{mck99} however points out that ambipolar diffusion might already have altered the mass-to-flux ratio observed by Crutcher.  Observations by \citet{bou01} yield similar conclusions to \citet{cru99}, although they caution that the measured magnetic field strength may be significantly lower than the true values due to the low volume filling fraction of dense cores.  \citet{nak98} argues that observed cloud cores cannot be magnetically supported.  A magnetically supported core should not have a column density much higher than its surroundings, and the core cannot maintain a large velocity dispersion inside it for its whole lifetime.  However, observed cloud cores generally have superthermal velocity dispersions, and column densities much higher than their surroundings, suggesting that most cloud cores are not magnetostatically supported.

Significant progress in the understanding of the roles of turbulence and MHD waves in supporting gravitationally unstable cloud cores has been made recently using 3D numerical simulations \citep[e.g.][for a review, see \citealt{mac04}]{vaz96,mac98,pad99,ost99}.  \citet*[hereafter Paper I]{kle00b}, using models computed with both ZEUS-3D and a smoothed particle hydrodynamics (SPH) code, conclude that hydrodynamical turbulence can prevent global collapse but not local collapse under typical molecular cloud conditions.  \citet*[hereafter Paper II]{hei01} include magnetic fields and conclude that local collapse cannot be prevented much longer than a global free-fall time by magnetized turbulence, in the absence of mean-field support.  However, they find that magnetic fields delay local collapse by decreasing local density enhancements via magnetic pressure.  Stars begin to form quickly when local density enhancements collapse.  This result favors the dynamical picture of molecular clouds being a transient feature in the interstellar medium \citep{bal99,elm00,har01} instead of existing for many free-fall times.

Another interesting result from the 256$^3$ resolution simulations of Papers I and II is that several cores begin to evolve into flattened (or disk-like) objects, which leads to the speculation that they were beginning to resolve the formation of protostellar cores.  Unfortunately, the resolution and number of disk-like cores were not high enough to perform a statistically meaningful study.  Therefore, these flattened cores were not discussed in Paper II.

Simulations suggest that supersonic turbulent flows, driven at large scale by supernovae and density waves, produce shock-compressed gas sheets or filaments that fragment into dense cores, and drive the observed supersonic turbulence in the clouds \citep{mac04}.  Differential rotation of galactic disks \citep{fle81} provides another plausible driving mechanism to maintain interstellar turbulence, especially at galactic radii with little massive star formation. The coupling from large-scale shear down to turbulent scales could be through the magnetorotational instability \citep[e.g.][]{bal98,sel99,kim03}.  Ionizing radiation \citep{mck89,vaz95,ber97,kri02}, and stellar winds from massive stars may also be important, but supernovae \citep[e.g.][]{nor96,avi00} probably provide most of the energy required to maintain interstellar turbulence as well as the turbulence within molecular clouds \citep{mac04}.

Recent numerical studies indicate that supersonic turbulence may play an important role in shaping the initial mass function (IMF) of the stellar population that forms in molecular clouds \citep{pad02}.  Observations of the IMF appear consistent with a general form having a power law in the high-mass wing and then flattening and turning over at the low mass end \citep[e.g.][and references therein]{kro02}.  This is an important constraint on star formation theories.  Interestingly, and perhaps significantly, the probability distribution function (PDF) of gas density in isothermal supersonic turbulence is well approximated by a log normal distribution.  \citet{pad97} and \citet{pad02} suggest that the IMF and density PDF in turbulent clouds are intimately related, and predict that the mass distribution of dense cores is controlled by the density, the Mach number, and velocity dispersion of the supersonic turbulence.  Previous simulations of molecular cloud collapse \citep[e.g.][]{pad01,kle01,bat03,gam03} indicate that the mass distribution of cores (or clumps) roughly resembles a log normal distribution, but with the high-mass wing following a power law approximately consistent with the Salpeter law.  However, these simulations either lack magnetic fields or do not have a sufficient number of gravitationally bound cores for accurate statistical analysis, because of low numerical resolution.  Higher resolution MHD turbulence simulations are required to verify the importance of super-Alfv\'{e}nic turbulent fragmentation to the stellar IMF.

In this paper, we report on results of our latest three-dimensional (3D), turbulent simulation of a magnetized molecular cloud, increasing the spatial resolution in each dimension by a factor of two relative to the simulations in Paper II, using the ZEUS-MP code.  The excellent parallel scalability of ZEUS-MP allows us to study MHD turbulent molecular clouds on a 512$^3$ grid, sufficient to study the global physical properties of cores and their formation process.  In \S2, we describe the techniques, assumptions, and parameters used in our simulations.  In \S3, we discuss the statistical properties of the super-Alfv\'{e}nic turbulent cloud before gravitation is turned on.  The global physical properties of self-gravitating cores and their time evolution are reported in \S4.  We also examine the turbulent fragmentation theory by comparing its predictions to our simulation results.  In \S5, we perform a resolution study and investigate the physical properties of some disk-like cores observed at later stages of our simulation.  We discuss our results and present conclusions in \S6.

\section{Simulations}

Our simulations are performed on 256 SGI/Cray Origin 2000 processors at the National Center for Supercomputing Applications using the ZEUS-MP code.  ZEUS-MP is a multi-physics, massively-parallel, message-passing code for astrophysical fluid dynamic simulations in 3D, developed by the Laboratory for Computational Astrophysics (lca.ncsa.uiuc.edu) at NCSA.  ZEUS-MP is a distributed memory version of the shared-memory code ZEUS-3D, which uses block domain decomposition to achieve scalable parallelism \citep{nor00}.  The code includes ideal hydrodynamics, ideal MHD, and self-gravity.  Self-gravity is implemented using the FFTW library \citep{fri98}.

Our scale-free MHD turbulent simulations are basically the same as model Eh1w in Paper II, but performed at resolutions up to 512$^3$.  All parameters are given in normalized units, where physical constants are scaled to unity.  The total mass in the box $M$ = 1 and the side length $L$ = 2.  The average density $\rho_o$ = 0.125.  The initial conditions of our simulation are identical to model Eh1w in Paper II.  In this model, the sound speed $c_s$ = 0.1.  The corresponding Jeans length $\lambda_{\rm J}$ = 0.5, and Jeans mass
\begin{equation}
{\rm M}_{\rm J} \equiv \rho_o \lambda_{\rm J}^3 = \left( \frac{\pi} G \right)^{3/2} \rho_o^{-1/2} c_s^3 = 0.0156.
\end{equation}                                               

Therefore, the mass in the box $M$ = 64 M$_{\rm J}$ and the length of the box $L$ = 4 $\lambda_{\rm J}$.  The magnetic field is initially uniform and along the z-direction with strength $B_o$ = 0.188.  The ratio $M/M_{\rm cr}$ = 8.3, which means our entire cloud is magnetically supercritical \citep[see e.g.][]{mou76}.  The ratio of thermal to magnetic pressure is 0.9 and the rms Mach number for this parameter set is $M_{\rm rms}$ = 10.  Isothermality is assumed throughout the simulation, which is justified as the cooling time is much shorter than the dynamical time inside high density molecular clouds.  No ambipolar diffusion is included in the calculation.

Periodic boundary conditions are applied in all three dimensions.  In effect we simulate a small portion of a larger cloud.  We do this for simplicity, to avoid addressing the structure of the non-isothermal boundaries of molecular clouds.  We do not expect that our results will be contradicted by unified models that include larger-scale flows in the interstellar medium.  In our analysis, we examine column-densities through one realization of our periodic cube, which should give statistically valid answers to the questions we ask for molecular clouds of finite size.

The procedure of setting up the turbulent flow is explained in \citet{mac99}.  We briefly review it here.  We assume that each component of the velocity perturbation is a Gaussian random field with flat power spectrum in the range 1 $\leq |k| \leq$ 2.  Random phases and amplitudes are generated in that spherical shell in Fourier space, and then transformed back into real space to generate each component of the driving velocity perturbation.  When all three components are obtained, the amplitude of the velocity is scaled to a desired initial root mean square velocity, which is unity in our case.  This velocity field is then normalized at each time step to ensure a constant kinetic energy input rate and added in as a perturbation to the evolved velocity field. The dynamical behavior of isothermal self-gravitating gas is scale free.  Equations (2) to (4) can be used to scale back to astrophysical units with a mass scale of the thermal Jeans mass, length scale given by the thermal Jeans length, and a time scale given by the free-fall timescale (see Paper I),
\begin{equation}
L = 0.89 {\rm pc} \left( \frac{c_s}{0.2 {\rm km}{\rm s}^{-1}} \right) \left( \frac{n}{10^4 {\rm cm}^{-3}} \right)^{-1/2}
\end{equation}                                               
\begin{equation}
M = 413 {\rm M_\sun} \left( \frac{c_s}{0.2 {\rm km}{\rm s}^{-1}} \right)^3 \left( \frac{n}{10^4 {\rm cm}^{-3}} \right)^{-1/2}
\end{equation}                                               
\begin{equation}
\tau_{\rm ff} = \left( \frac{3 \pi}{32 G \rho} \right)^{1/2} = 0.34 {\rm Myr} \left( \frac{n}{10^4 {\rm cm}^{-3}} \right)^{-1/2}
\end{equation}                                               

Therefore, if we choose a high-density scaling such as the BN region in Orion, in which $c_s$ = 0.2 km s$^{-1}$ and the number density $n$ = 10$^5$ cm$^{-3}$, the region in the simulation box is 0.28 pc in size.  The total mass in this region is 130.6 M$_\sun$ and one system time unit, which is 0.65 free-fall time, is 0.07 Myr.

A total of 110,000 CPU hours were required for the 512$^3$ zone simulation to run to 4.325 time units, or $\sim$ 2.82 $\tau_{\rm ff}$.  In our case, $\tau_{\rm ff}$ = 1.53 system time units.  At the beginning of the simulation, no gravitational forces are applied.  Turbulence is allowed to develop from the relatively smooth initial conditions until $t$ = 2.0 time units.  Gravitation is then turned on.  Density fluctuations generated by the supersonic turbulence in converging and interacting shock fronts that locally exceed the Jeans limit begin to contract.  In the 256$^3$ simulations reported in Paper II, there were only a dozen gravitationally collapsing cores, which were under-resolved and insufficient for statistical analysis.  In our 512$^3$ simulation, there are 15 cores that satisfy our definition of gravitationally bound at $t$ = 2.0 when gravity is turned on.  By the end of the simulation at $t$ = 4.325, we have 83 gravitationally bound cores.  The simulation was terminated then because too many cells violate the Jeans condition \citep{tru97} which dictates that the local Jeans length must be resolved by several zones.  Also, the densities of some cores became so high that the isothermal assumption would be invalid.  Henceforth, we adopt the same time convention as in Paper II, where the time $t$ = 0 is defined as the time that gravity is turned on.   According to this definition, the simulation is terminated at $t$ = 2.325 system time, or 1.52 $\tau_{\rm ff}$.  We will use free-fall time as the time unit for the rest of the paper.

\section{Statistical Properties}

\subsection{Power Spectra}

\citet{kol41} heuristically derived a simple scaling for the 1D energy spectrum in incompressible turbulence, $E(k) \propto k^{-5/3}$, where $k$ is the magnitude of the wavenumber.  Astrophysical fluids are magnetized and highly compressible.  Dynamically important magnetic fields and strong shocks should interfere with eddy motions, which in turn will affect the final velocity power spectrum.  However, observations and some simulations \citep[e.g. Paper I; ][]{laz02,cho03} suggest that MHD compressible turbulence also closely follows the Kolmogorov spectrum.  The theory of \citet{gol95} points out that it is the perpendicular motion of sub-Alfv\'{e}nic incompressible turbulence that behaves as Kolmogorov, $E(k) \propto k^{-5/3}$.  Some other MHD turbulence simulations suggest that the scaling law for MHD compressible turbulence is not exactly Kolmogorov.  \citet{bol02} reported that the power spectrum of the solenoidal component of the velocity field in a super-Alfv\'{e}nic compressible turbulence is $E(k) \sim k^{-1.74}$, steeper than the Kolmogorov spectrum.  In the latest simulations by \citet{ves03} on 256$^3$ and 512$^3$ grids, the slope of the power spectra varies from Kolmogorov to even larger than the  $E(k) \sim k^{-2}$ Burgers' model of shock-dominated turbulence \citep{bur74} for models ranging from sub-Alfv\'{e}nic to super-Alfv\'{e}nic.  The measurement of the slopes in their simulations might be affected by the short (or nonexistent!) inertial range of their power spectra, however.

Figure 1a shows the power spectra of density, velocity, and the energies of our turbulent cube at $t$ = 0, after the turbulence has fully developed, but before gravity has been turned on.  The long dashed line is the driving spectrum.  The power spectrum of velocity averaged over all directions is basically a Kolmogorov power spectrum (the thin straight line in Figure 1a).  The inertial range of the velocity power spectrum spans more than an order of magnitude of the wavenumber (2 $\leq |k| \leq$ 30).  The power spectrum of the magnetic field shows a similar result.  The density spectrum has a shallower slope with index $\sim$ -1.15.  Both the kinetic and potential energies have shallower slopes because of the modulation by the density spectrum.  Figure 1b is a compensated plot of the velocity power spectrum multiplied by $k^{5/3}$ that highlights the excellent fit of the Kolmogorov power spectrum to the velocity power spectrum.  We also plot the compensated velocity power spectrum for the 256$^3$ resolution simulation for comparison.  The inertial range of the turbulent flow in this lower resolution simulation is shorter but still close to the Kolmogorov power spectrum.

\subsection{Density PDF and Core Mass Spectrum}

The PDF provides important and complementary information to the power spectrum.  The PDF of isothermal turbulent flows can be approximated by a log normal function \citep[e.g.][]{vaz94,pad97,sca98,nor99}.  With a log normal PDF of mass density, most of the material concentrates in a small fraction of the total volume in the simulation.  Figure 2 shows the best log normal fit to the PDFs of all the cells in the 256$^3$ and 512$^3$ simulations at $t$ = 0.  The standard deviation of the best fit is 1.46 $\pm$ 0.06 and 1.42 $\pm$ 0.03, as defined in equation (6) in \citet{pad02}.  Using equation (9) in \citet{pad02}, $\sigma^2$ = ln(1 + $M_{\rm A,rms}^2 \eta^2$), we calculate $\eta$ to be 0.41 and 0.38, respectively, a little smaller than the value $\eta \sim$ 0.5 reported in \citet{pad02} from numerical experiments.

Figure 3 shows the projected density (column density) images of the simulation, in logarithmic scale, along the x-, y-, and z-direction, which shows the clumpiness expected from a log normal PDF.  The first row of plots are at $t$ = 0, just before gravitation is switched on.  The second and third rows are at $t$ = 0.65 $\tau_{\rm ff}$ and $t$ = 1.31 $\tau_{\rm ff}$, respectively.  At $t$ = 0 (Figures 3a-c), filamentary structures dominate the view and we see some higher density nodes on the filaments.  Later (Figures 3d-f), gravity breaks the filaments into pieces and higher density clumps slowly increase their mass by accretion.  The peak density at this moment is about 1000 times the average density.  At $t = 1.31 \tau_{\rm ff}$ (Figures 3g-i), many cores with high column-densities have formed.  The peak density is about four orders of magnitude above the average density and the total mass in cores is less than 9\% of the total mass in the box.  Qualitatively, this picture does not change during the subsequent simulation time (up to $t = 1.52 \tau_{\rm ff}$), except that more cores form and their peak densities grow.  Most of the volume in the simulation corresponds to low density voids.  The clumpiness can also be seen in Figure 4 which shows a volume rendering of the logarithmic density and the magnetic pressure of the turbulent box in 3D at $t$ = 1.31 $\tau_{\rm ff}$.  An animation of the evolution of density and magnetic pressure of the 512$^3$ simulation is included in the electronic version of this paper.

\citet{pad02} propose that the stellar IMF is a direct result of turbulent fragmentation of molecular clouds.  As shown in Figure 3, the gas is compressed into 1D filaments and 2D sheets by supersonic turbulence.  Therefore, the core mass distribution will be dependent on the jump conditions for isothermal shocks in a magnetized gas.  The typical size of a dense core is comparable to the thickness of the post shock gas.  Based on this, \citet{pad02} calculate by taking into account the scale-dependent Mach number that the mass distribution of cores formed inside a supersonic MHD turbulent cloud is
\begin{equation}
N(m) d log(m) \propto m^{-3/(4-\beta)} d log(m)
\end{equation}
where $\beta$ is the spectral power index.  With our 512$^3$ simulation, we have sufficient resolution to check the turbulent fragmentation prediction.  We proceed as follows.

The clumpy structure inside the turbulent molecular cloud in our simulation is determined using the algorithm CLUMPFIND described by \citet{wil94}.  The algorithm is also described in Paper I.  A ``core'' is defined as a gravitationally bound region.  We define it to consist of a set of connected zones with average density exceeding the mean density expected for isothermal shocks, $\rho > M_{\rm rms}^2 \rho_{\rm o}$, potential energy exceeding kinetic energy, $|E_{pot}^{cell}| > E_{kin}^{cell}$, and core mass exceeding the local Jeans mass, $M_{core} > M_{\rm J}(\rho)$.  We use logarithmic density contours in CLUMPFIND to get a wide enough density range, as in Paper I.  We believe the core definition chosen here is more rigorous than the simple gravitationally unstable overdensity criterion of \citet{pad01}.  However, our definition identifies fewer cores.  This definition of core will be applied to all results on cores in this paper.  We also define ``clumps'' to be overdense regions that do not satisfy the other criteria.  We choose the overdensity threshold for clumps somewhat arbitrarily to be $\rho > 0.1 M_{\rm rms}^2 \rho_o$

In Figure 5, we plot the mass spectra of cores (dot-dashed line) and clumps (dashed line) in the 256$^3$ simulation and the spectrum of cores (solid line) in 512$^3$ simulation at $t$ = 0, normalized by the Jeans mass $M_{\rm J}(\rho_o)$.  There are only 11 cores in the 256$^3$ simulation.  Therefore, we also plot the clump spectrum for comparison.  Many of these clumps could be just density fluctuation created by the isothermal shocks that will be destroyed later.  The 256$^3$ clump spectrum and 512$^3$ core spectrum both show a well-defined ``universal'' IMF \citep[see the latest review by][]{kro02}.  From Figures 1a and 1b, the velocity power spectral index $\beta \approx$ -5/3.  Substituting $\beta$ into equation (5), we expect the slope of the high-mass wing of the core mass spectrum will be -1.29.  This spectral index is plotted as thin line in Figure 5 and is consistent with the high-density wings of all the spectra, as predicted by the turbulent fragmentation.  Even the 256$^3$ core spectrum appears consistent with the prediction.  Note that the number of cores in Figure 5 is still too few to have an accurate fitting unless we relax the definitions of the cores to clumps.

\section{Cloud Core Evolution}

\subsection{Delay of Core Collapse}

Paper I and Paper II concluded that local collapse could not be prevented even when the turbulent velocity field carries enough energy to balance the gravitational contraction on global scales.  Paper II found that the comparison between 128$^3$ and 256$^3$ MHD results did not show full convergence, with increasing resolution delaying local collapse (see Figure 6 in Paper II).  This occurs because increasing the resolution produces thinner shocks with peak densities closer to the theoretical value.  The resulting deeper potential wells accrete more mass and collapse more quickly.  At the same time, the simulation can better follow the short-wavelength MHD Alfv\'{e}n waves, which could create an isotropic pressure to counteract the gravitational collapse.  From the 256$^3$ simulations in Paper II, it appears that the MHD waves are able to delay the collapse but not to prevent it.  In our new 512$^3$ simulation, this imbalance between gravitation and Alfv\'{e}n wave pressure is verified.  Figure 6 combines our new results with the previous 64$^3$, 128$^3$, and 256$^3$ results.  We plot the collapsed mass $M_*$, defined as the sum of the core masses, versus time.  Collapse still occurs in all cases.  Comparison between the 256$^3$ and 512$^3$ models suggests that the 256$^3$ result of Paper II is, in fact, converged.  Remaining discrepancies are produced by the chaotic nature of the flow, as was shown by comparisons of different realizations of the same resolution model in Paper I.

\subsection{Core Formation and Evolution}

In the models, we find that self-gravitating cores form out of initially overdense clumps created by supersonic turbulence.  We also find that dense clumps can be destroyed by passing shocks as well as by the merging of clumps to form larger and denser objects.  The formation and destruction of cores is thus a complex dynamical process.  The core formation and destruction process is shown in Figure 7 in terms of number of cores formed as a function of time.  Instead of monotonically increasing, the core number drops sometimes because of merging or destruction.

Figures 8 to 10 show the evolution of the physical properties of cores after gravitation is turned on and before the Jeans criterion \citet{tru97} is violated.

Figure 8a and 8b show the ratio of core mass to the mass that the local magnetic field can support, $M/M_{\rm cr}$.  The ratio
\begin{equation}
\frac{M}{M_{\rm cr}}=\frac{M/\Phi}{(M/\Phi)_{\rm cr}}=\frac{N}{F B_{\rm los}}{\frac{\sqrt{G}}{c_{\Phi}}},
\end{equation}
where $N$ is the column density; the geometrical correction factor $F = 2$ for a uniform sphere and 3 for an isothermal disk.  $B_{\rm los}$ is the line of sight magnetic field strength; $G$ is the gravitational constant; and $c_{\Phi}$ $\sim$ 0.12 for a spherical cloud or 0.16 for an isothermal disk \citep{bou01}.  In these two plots, we choose $F = 3$ and $c_{\Phi} = 0.16$ for the extreme case.  The ratios exceed an order of magnitude at all time.  If we assumed spherical cloud geometry, the ratios would be even higher.  Therefore, all the cores are magnetically supercritical during the gravitationally collapsing phase, despite the action of gravitational fragmentation, which could be expected to reduce the mass to flux ratio \citep{mes56}.

Figures 8c and 8d show the relationship between the central magnetic field strength $B_c$ and the central core density $\rho_c$ at $t$ = 0 and $t$ = 0.65 $\tau_{\rm ff}$ respectively.  The power law of index 0.5 is also plotted.  Interpretations of $B_c \propto \rho_c^{1/2}$ can be that: 1) cores are magnetically subcritical \citep[e.g.][]{mou91,bas94,bas95}, 2) cores are magnetically supercritical but the Alfv\'{e}nic speed is about constant ($\sim$ 1) for the cores \citep[e.g.][]{ber92,cru99}, 3) cores are magnetically supercritical and dense cores tend to accrete mass along magnetic field lines and reduce their magnetic flux to mass ratio efficiently, even in the absence of ambipolar diffusion \citep{pan99}.  The cores in the simulations are all magnetically supercritical and there is no ambipolar diffusion implemented in ZEUS-MP.  Therefore, Figure 8 could be a result of case 2) or 3).  Magnetically subcritical cores are no longer a unique explanation to the $B_c \propto \rho_c^{1/2}$ relation.

In Figures 9a and 9b, we plot the specific angular momentum, $j$, of each core against the radius, R, of the core at $t$ = 0 and 0.65 $\tau_{\rm ff}$.  The radius $R$ is determined from the maximum dimension of the core determined using CLUMPFIND.  The majority of cores obey the power law relationship, $j \propto R^p$, with index $p$ = 3/2, which is plotted for comparison.  From observations, \citet{goo93} concludes that $j$ scales roughly as $R^{3/2}$ based on the velocity gradients of 29 cores in dark clouds.  They interpreted this relationship between specific angular momentum and core size to imply that cores are in approximate "virial equilibrium".  With the isothermal equation of state and relatively weak magnetic field support in our simulation, the equilibrium in cores is basically between gravitational force and rotational angular momentum, producing rotationally supported cores.  In their study of rotational properties of centrally condensed, turbulent, molecular cloud cores, \citet{bur00} conclude that the specific angular momentum is related to the radius by $j \propto \Omega R^2$, where $\Omega$ is the angular speed of the core.  Therefore, $j = \Omega R^2\propto R^{3/2}$ implies $\Omega \propto R^{-1/2}$.  This agrees reasonably well with observations that suggest $\Omega$ scales roughly $R^{-0.4}$ \citep{goo93,bar98}.  However, we cannot conclude just from Figure 9 that the cores are in Keplerian rotation, $\Omega(r) \propto r^{-0.5}$, or in angular momentum conserved rotation, $\Omega(r) \propto r^{-1}$.  In addition, there is still a slight scattering of data in Figure 9.  The rotation could indeed be Keplerian at the inner parts of the core but angular momentum conserved in the outer parts.

In Figure 10a, we plot the relationship between mean density and radius of each core at $t$ = 0 (circle) and $t$ = 0.65 $\tau_{\rm ff}$ (triangle), where mean density $<\rho> = M_{core} / V_{core}$, and $V_{core}$ is the volume of the core.  The result agrees with that of \citet{bal02}  (see their Figure 9) that mean density is basically independent of radius.  In Figure 10b, we plot the mass to radius relationship of each core.  A power law relationship is seen, with $M_{core} \propto R^{2.75}$ from $t$ = 0.  Figures 10a and 10b would seem to be inconsistent because we would expect $M_{core} \propto R^3$ if cores are spherical.  If we have $V_{core} \propto R^{2.75}$ from Figure 10, that means either that the geometry of the cores lies between a sphere and a disk, or that the cores are a mixture of prolate, triaxial, and oblate objects, as data points are scattered around the straight line.  We investigate the shapes and orientation of cores in \S4.4, and confirm this conjecture.  The correlations seen in Figures 8 to 10 remain valid for the whole simulation even after \citet{tru97} criterion is violated.  They are also found at lower significance in the 256$^3$ resolution simulation.

The number of cores found with CLUMPFIND from $t$ = 0.0 to $t$ = 2.25 basically increases with time, as shown in Figure 7.  Some cores are destroyed by the supersonic turbulence in the simulation.  At the same time, some other cores actually merge to form more massive cores through gravitation \citep[e.g.][]{bon97,kle00a,bon01a,bon01b}.  Figure 11 shows the merging of two cores, or, to be more specific, the accretion of a small clump by a massive clump.  The three columns of subplots in Figure 11 show views along three different axes of the merging sequence.  The time interval between two rows is 0.05 $\tau_{\rm ff}$.  As the result of core mergers and accretion of surrounding material, the final cores possess much larger angular momentum than the initial clumps.  This leads to the formation of disk-like cores during the simulation.  This phenomenon is entirely absent in isolated star formation simulations, which consider only a single, gravitationally bound core.

\subsection{Core Mass Spectrum Evolution}

It is interesting to examine the effect of core mergers on the core mass spectrum.  In Figure 12, we plot the evolution of the core mass spectrum from $t$ = 0 to the end of the simulation.  In each panel, we plot a power-law with slope -1.29 for reference.  As discussed in \S3.2, at $t$ = 0, the core mass spectrum resembles closely a log normal distribution and the slope of the high-mass wing is consistent with the power law with index predicted by turbulent fragmentation theory \citep{pad02}.  There are some fluctuations of the core mass spectra at later time, but the slopes of the high-mass wing remain about the same until $t \sim 1 \tau_{\rm ff}$.  The slope becomes shallower at yet later times, but we do not have too much confidence in the correctness of the core mass spectra after $t = 0.65 \tau_{\rm ff}$, as the centers of the cores become unresolved.  The hydrodynamical simulation by \citet{kle01}, using SPH and sink particles, also shows shallower slope at the later stages of the simulation, which he attributes to the coalescence of cores, but again, fragmentation in the centers of cores is suppressed, in this case by the use of sink particles.  Even if the trend of increasingly shallower slope in core mass spectra is qualitatively correct in our simulation, the rate of slope increase may be over-estimated.  Our assumed periodic boundary condition means the cores have no place to disperse except merging in the later stage.  The finite resolution in the simulation prevents cores from collapsing as far as real cores would, so that the simulated cores have larger cross sections for merging (Paper I).  Nevertheless, for a dense cluster, the core merging rate may still be high.  Recent millimeter wave observations of cloud cores also reveal mass spectra of clumps similar to the stellar IMF \citep[e.g][]{mot98,tes98}, with slope of the high-mass wing of $\sim$ -1.1 to -1.5, closer to Salpeter -1.35 than that derived for gaseous clumps of -0.5 \citep{bli93}, using the form of IMF definition in equation (5).  Our simulations are in good agreement with these studies.
 
Because of the higher resolution of our 512$^3$ simulation, we have many more cores to analyze than in \citet[see their Figure 4]{gam03} and we are able to demonstrate a core mass distribution similar to the general IMF.  Note that not all material within a core will finally end up in the protostar.  In \S5, we will find out that more than half of the core mass resides outside the central region.  Even material inside the central region may be evaporated or ejected away during star formation.  In addition, the cores that we observed in the simulation may further fragment into binaries or even multiple systems if the resolution of the simulation can be increased further.  The \citet{tru97} criterion is eventually violated at the centers of our cores (see \S5) so we cannot follow this evolution.  We note that, despite the emphasis on artificial fragmentation because of the particular test problem chosen by \citet{tru97}, regions that are underresolved by their criterion may also show too little fragmentation if the geometry is less pathological.  Competitive coagulation \citep[e.g][]{sil79,lej86,mur96,bon01a,bon01b} may reshape the final IMF when gravitation becomes more important.  Therefore, we cannot reach a definite conclusion that the final stellar IMF is similar to the core mass spectrum.

\subsection{Core shapes and Orientation}

\citet{gam03} recently studied gravitationally collapsed clumps in decaying MHD turbulence simulations on 256$^3$ grids with different magnetic field strengths.  In our 512$^3$ driven turbulence simulation, $v_A/c_s$ = 1.5, which is close to the value for their run C.  Using the same definition and symbols of principal axis lengths and ratios as in \citet{gam03}, we calculated the principal axes of the cores in our simulation at $t = 0, 0.65 \tau_{\rm ff}$, and 1.31 $\tau_{\rm ff}$  and plotted the axis ratios in Figure 13.  It is useful to divide the lower triangular part of the figure into 3 regions: a prolate region of $b/a$ = 0 to 0.33, a triaxial region of $b/a$ = 0.33 to 0.66, and an oblate region of $b/a$ = 0.66 to 1.  The result from our higher resolution, driven run basically resembles that of \citet[see their Figure 12]{gam03}.  The majority of cores are prolate or triaxial in shape, even near the end of our simulation.  We also see a small number of cores that become oblate late in the simulation, with large axis ratios.  This implies that some cores become disk-like.  Figure 14 shows the angles between the shortest core axis and the density weighted mean magnetic field in cores at $t = 0.65 \tau_{\rm ff}$ (dot-dash line) and 1.31 $\tau_{\rm ff}$ (solid line).  We see a slight correlation in the angles between the shortest core axis and the density weighted local mean magnetic field at $t = 0.65 \tau_{\rm ff}$ and an even stronger correlation at 1.31 $\tau_{\rm ff}$, which is different from the result of \citet{gam03}.  This difference could possibly be a result of higher resolution in our simulation.

\section{Disk-like Cores}

In Figure 13, we see a relatively small number of cores inside the oblate region.  The cores are too small to study at $t$ = 0 but large enough to study at $t = 1.31 \tau_{\rm ff}$.  Most of these cores have a well defined disk-like appearance that could provide some hints about protostellar disk formation.  Before we begin the discussion of disk-like core, we must emphasize that for some cores, the density of the inner cells grows so high that that the local Jeans length is no longer resolved in their centers, violating the \citet{tru97} criterion.  Unfortunately, these cores are also the largest whose physical properties can be studied with reasonable statistical significance.  Other models using SPH and sink-particles also find disk-like cores in the later stages \citep[e.g.][]{bat03}.  Although our cores are under-resolved at $t = 1.31 \tau_{\rm ff}$, the external disks of cores may still provide some useful information on the dynamics of core accretion.  With the clear realization that even higher resolution will ultimately be needed to properly address the structure of these objects, we proceed with an analysis of the data we have in hand, bearing in mind that our 512$^3$ model is the highest resolution driven, self-gravitating MHD simulation currently available.

\subsection{General Properties of disk-like Cores}

We select 12 cores inside or near the oblate region in Figure 13 with well defined disk-like appearance and tabulate their mass and diameters in Table 1.  The cores selected are shown in the surface density map (Figure 15).  Core numbering is arbitrary.  The orientation of the disk-like cores is uncorrelated to the 3D grid in the simulation despite even though they accidentally appear correlated in Figure 15.  We have examined the orientation of the shortest axis of each disk-like core and they are basically random with respect to the 3D grid.  The corresponding masses and radii using the high-density scaling mentioned in \S2 are also shown for easy comparison with observations.  The masses of cores range from 0.1 to 2.5 M$_\sun$ and the diameters of the disks are a few thousand AU.  The accretion rates of the cores will be discussed later in this section.  The 12 cores can be roughly categorized in two groups: (1) a cluster of seven cores that remain close to each other in a region with higher mean magnetic field, and (2) five cores that are basically isolated and with higher spatial velocity (see Table 1).  Cores in category 1 have higher mass, central density, magnetic pressure, and accretion rates compared to cores in category 2.  The mean magnetic field is an order of magnitude higher inside the cluster.  The size of both grouped and isolated cores appears similar.  By looking at these two categories of disk-like cores, we can see some manifestation of mass segregation in that low-mass cores are preferentially found at large radii from the cluster whereas massive cores sink towards the center \citep[e.g.][]{bon98}.

The mass and sizes of the 12 disk-like cores shown in Table 1, using high-density scaling, are in the range of the recent observations of young stellar objects that possess disk-like envelopes or circumstellar disks \citep[e.g.][]{zha98,fue01,hog01,jay01,wis01}.  Most of these observations suggest that the sizes of the circumstellar disks of many young stellar objects are just a few hundred AU and the sizes of the disk-like envelopes are a few thousand AU.  In addition, most of the studied protostars inside the disk-like cores are low or medium mass stars of mass $\leq$ 1 M$_\sun$.  The estimated mass of the circumstellar disk is only a small fraction of the central star.  The more massive objects we find in our simulation are better thought of as disk-like envelopes rather than true protostellar disks.

\subsection{Resolution study of disk-like Cores}

In order to better understand how well the cores are being resolved in our simulation, we compare them with typical cores formed in other lower resolution simulations.  In Figure 16, we show images of the projected column density of four cores from four simulations at the time when the total mass inside cores is $\sim$ 10\% of the total box mass.  In the 64$^3$ run (first row of Figure 16), there are four cores found by CLUMPFIND and we cannot identify any disk-like core (axis ratio $>$ 2:1).  In the 128$^3$ run, there are 24 cores found, and still no disk-like core is identified (second row in Figure 16).  In the 256$^3$ run, there are 38 cores determined and four are disk-like (third row in Figure 16).  In the 512$^3$ run, there are 67 cores and twelve are disk-like.  In the last row of Figure 16, we show core 1 from the 512$^3$ run.  Note that the coordinates shown in Figure 16 are grid zones, not spatial distance.  The spatial size of the core in 64$^3$ is the largest but the diameters of cores in 128$^3$, 256$^3$, and 512$^3$ are about the same, indicating the convergence of core size with increasing simulation resolution.  We can also see that the disk-like appearance of cores becomes more pronounced with higher resolution.  Figure 17 shows the surface density profiles of the cores in Figure 16.  The curves are offset for ease of comparison.  We can see a converging density profile of the core with almost uniform density at the center and a power law distribution of the disk/envelope with higher resolution, as predicted by other single cloud core simulations.  Core 1 has the largest spatial diameter and is selected for further study below.

\subsection{Physical Properties of a Disk-like Core}

In Figure 18, we show close-up column density images of core 1 projected along the x-, y-, and z-axes.  At the end of the simulation, core 1 seems to be an isolated core as it is far from other cores.  The closest neighbor is about half a box size ($>$ 0.14 pc) away from it.  In fact, core 1 passes through a dense region at $t \sim 1 \tau_{\rm ff}$, interacts with several cores, and accretes a large amount of material.  The disk only appears clearly after this; the core remains isolated thereafter.   Gravitational interaction among protostars in nascent star clusters is argued to tidally truncate or even disrupt accretion disks \citep[e.g.][]{cla91,hal96,sca01,kro01}.  The disk-like cores in this simulation suggest the alternative that disks could survive at least some interactions.  Core 1 is not massive ($\sim$ 0.457 M$_\sun$) but the extent of the accretion disk is the largest ($>$ 5000 AU) of all the cores in the simulation.  In the z-direction plot of Figure 18, we can see some evidence of spiral structure in the disk.

In Figures 19a-c, the radial profiles of energies, velocities, and surface densities of core 1 are shown.  All the physical quantities in the radial direction are the average values of zones binned at the same radius from the center of mass.  We can see that the magnetic energy remains relatively small compared to the kinetic energy (Figure 19a).  The radial velocity (dashed line), $v_{rad}$, of the disk shown in Figure 19b shows a typical picture of angular momentum transfer inside a viscous disk, in which friction tries to spin up the outer part of the disk and spin down the inner part \citep{lyn74}.  However, the largest source of viscosity is likely numerical in our model, so we cannot draw any quantitative conclusions from this figure.

From the choice of sound speed in the simulation, the unit velocity corresponds to 2 km s$^{-1}$.  In Figure 19b, the radial infall rate is $\leq$ 0.07 in model units, corresponding to $\leq$ 0.14 km s$^{-1}$, on the order of what is observed or modeled \citep{wil99,mye00}.  Error bars that plot variances of radial and rotational velocities estimated at each radius are also shown.  The disk is basically in a rotationally supported state.  Note that the inward or outward radial velocity of the disk is not the velocity used in estimating the accretion rate of material to the core in Table 1, because, as we will show later in this section, material mostly accretes onto the core in the direction perpendicular to the disk after the disk is well formed.

The dot-dash line in Figure 19b is the theoretical Keplerian rotation curve ($v_{rot} \propto r^{-0.5}$).  The actual core rotational curve (solid line), $v_{rot}$, shows typical Keplerian rotation in most parts of the disk.  The angular momentum conserved rotational curve (dotted line) is also shown for comparison.  The rotational speed of the core 1 is $\sim$ 0.25--0.35 within the radius of 7 zones, which corresponds to $\sim$ 0.13--0.18 rad Kyr$^{-1}$ in angular velocity.  It takes about 40,000 yr for the core to complete one full rotation.  Figure 19c shows the surface density profile (solid line) along with the best fitting power law of surface density  $\propto r^{-2.4}$ (dash line).

Observationally, most of the information about true protostellar disks is obtained by using the position-velocity (PV) diagram \citep{ric91}.  Note that the information on rotation and infall obtained from PV diagrams is subject to large uncertainties if the source is poorly resolved.  Figure 20 shows the PV diagram of core 1, observed in the x-direction along the y-axis (see Figure 19a).  The contours in the PV diagram are logarithmic in column density, which is proportional to the observed intensity if the disk is optically thin and isothermal.  In Figure 20a, the PV diagram is shown at full resolution.  The relative position ($\Delta$Y) in zones is plotted against the LSR velocity in model units.  The shifted appearance of velocity in the PV diagram is the result of disk rotation.  For qualitative comparison to observations, we convolve the map with a circular Gaussian beam having FWHM of three grid zones (Figure 20b).  The dashed curve shows the distribution of the rotation velocity proportional to $r^{-1}$, which is characteristic of angular momentum conserving rotation.  The solid curve shows the distribution of the rotation velocity proportional to $r^{-0.5}$, which is characteristic of Keplerian rotation.  We can see that Keplerian rotation better describes most parts of the disk in this core, although the poor resolution of the PV diagram limits the strength of this conclusion.  If we use this PV diagram to estimate the point mass of the central star, we would conclude that the central mass of core 1 is about 0.21 M$_{\sun}$, using the high-density scaling, which is the mass enclosed by the region of radius $<$ 3.4 zones at the center of this core.  We find that the mass of the disk outside this central region is more than 53\% of the total core mass (Table 1).  An example of an object with properties similar to core 1, in high-density scaling, is IRAS 05413-0104 \citep{wis01}.  IRAS 05413-0104 is a flattened disk-like core of average volume density 10$^5$ cm$^{-3}$, with an envelope of mass $\sim$ 0.2 M$_{\sun}$, and a disk with diameter of $\sim$ 12,000 AU, roughly twice the size of core 1.

In Figure 21, we show the velocity and magnetic field components perpendicular and parallel to the disk for core 1.  The velocity field on the disk plane clearly shows rotation (Figure 21b).  The edge-on view of the velocity (Figure 21a) shows that material above and below the disk is mostly falling onto the core along the magnetic field lines, as expected when magnetic field is present.

The accretion rates of all cores at $t = 1.31 \tau_{\rm ff}$ are listed in Table 1.  The accretion rate is calculated by integrating the material falling onto the whole area of the disk above and below.  From Table 1, the accretion rates of the 12 cores are in the range of $\sim 10^{-4}$--$10^{-5}$\ M$_{\sun}$ yr$^{-1}$.  If we assume the accretion rate is constant, the ages of the cores are about 15,000 to 50,000 years old, which is only $\sim 0.14$--0.46 $\tau_{\rm ff}$.  Therefore, the accretion rates of cores must have dramatically increased near $t = 1.31 \tau_{\rm ff}$, as expected from a core collapse scenario.  The rate of accretion onto core 1 is $\sim$ 9.72 $\times 10^{-6}$ M$_{\sun}$ yr$^{-1}$.  On average, the cores inside the cluster have accretion rate more than twice of those outside.  The accretion rates of all 12 cores are basically at the high-end of the range of observed envelope infall rates for protostellar cores \citep{ost98}.

In \S4.4, we noted a weak correlation between the core minor axis and the local magnetic field direction.  Here in Figure 21c, the vertical component of magnetic field lines does show a bipolar appearance in core 1, but the magnetic field is bent, due to the movement of the core relative to the ambient material.  The magnetic field component along the disk also shows rotating features, which indicates that the magnetic field is frozen and dragging along with the rotating disk (Figure 21d).  This magnetic structure is common for all the disk-like cores.  In Figure 22, we show the 3D structure of the magnetic field in core 6, which is helical as a result of core rotation.  The angular momenta of cores are expected to be transferred to the ambient material by such helical Alfv\'{e}n waves, which in return brake the rotation of the cores \citep[e.g.][]{mou79,mou80,bas94}.

\section{Conclusion and Discussion}

Molecular clouds are magnetized and in a state of turbulent motion.  It remains an open question whether the turbulence is sub-, trans-, or super-Alfv\'{e}nic \citep[e.g.][]{mac04}.  Incompressible hydrodynamic turbulence has been thoroughly studied, but compressible MHD turbulence is still a relatively young area, especially as applied to astrophysical problems.  In the last decade, numerical studies of compressible turbulence in molecular clouds have revealed important results that may directly relate to the early stages of star formation.  Such simulations still suffer from either lack of magnetic field, lack of self-gravity, or insufficient resolution.  In this paper, we present the results of our latest studies on self-gravitating core formation in a super-Alfv\'{e}nic turbulent molecular cloud.

A statistical analysis of the simulation after the turbulence has fully developed shows that the power spectra of velocity and magnetic field basically follow Kolmogorov scaling, even though the system is compressible and magnetized.  This result is consistent with some other recent high resolution studies on super-Alfv\'{e}nic turbulence \citep[e.g.][]{mul00,cho03}.  The power spectra for density, kinetic, and potential energies show shallower power-law indexes.

We calculated the core mass spectrum in our high-resolution simulation including both MHD and self-gravity, and followed its evolution.  At least at early times when the model is completely resolved, our core mass spectrum is consistent with two main predictions of the turbulent fragmentation theory advanced by \citet{pad02}: that the slope of the velocity spectrum defines the slope of the high-mass wing of the core mass spectrum, and that the low mass end flattens and turns over.  \citet{pad02} propose that this could account for the form of the stellar IMF.  At later times in our model, the slope of the high-mass wing becomes shallower, possibly because of core coalescence or lack of resolution of fragmentation in central regions of cores.  Our core encounter rate may well be overestimated because of the choice of periodic boundary conditions and grid resolution.

The simulation that we have presented in this paper is based on initial conditions that are both favorable and hostile to the formation of cores.  In the case simulated, the molecular gas is magnetically supercritical with $M/M_{\rm cr}$ = 8.3.  We expect {\it a priori} that gravitational collapse is unavoidable and will occur eventually.  At the same time, the system is subjected to a strong turbulent driving force producing flows with rms thermal Mach number $M_{\rm rms}$ = 10.  The turbulence is strong enough to formally support the entire region, so it could prevent or delay the collapse of cores or even destroy some cores in the process of collapsing.  In Paper II, it was shown that in a simulation on a 256$^3$ grid, with the same initial conditions, the magnetic and turbulent support could not prevent local collapse of cores in the molecular cloud.  Cores form almost immediately after the gravitational force is turned on.  One of the motivations for our higher resolution simulation---twice the spatial resolution and eight times the mass resolution of the 256$^3$ model of Paper II---is to determine whether stronger Alfv\'{e}n wave support may occur when shorter wavelengths are resolved in a simulation using the same initial conditions.  We find that gravitational collapse still wins over magnetic and turbulent support.  The 512$^3$ simulation lasts about 2/3 of a full free-fall time, after the gravitation is turned on, before the first cell in the computation violates the Jeans criterion \citet{tru97} that the local Jeans length be resolved, a factor of four longer than the 256$^3$ simulation.  A resolution study shows convergent trends with increasing simulation resolution of both the fraction of the box mass in bound cores (Fig. 5) and the core size (Fig. 17).

Our simulation began with the entire box magnetically supercritical.  However, it is often argued that gravitational fragmentation will produce smaller and smaller regions that eventually become subcritical \citep{mes56}.  We found that gravitationally bound cores formed and collapsed until they were rotationally supported (Fig. 9), while still having masses at least an order of magnitude over the critical mass for magnetic support (Figs. 8a-b).  We further found that the central magnetic field strength depends on the central density as $B_c \propto \rho_c^{1/2}$, suggesting that observations of such relations do not necessarily point to subcritical cores.

Most of the cores appear to be prolate or triaxial in shape, consistent with simulation results from \citet{gam03}.  However the correlation of the shortest axis to the density-weighted local magnetic field of the cores is found to be stronger in our simulation.

In our simulation, we observe interaction between cores during close encounters.  The continuing accretion of large amounts of material, and merging of cores results in cores with high specific angular momentum that helps the formation of disk-like structure.  The final outcome is highly unpredictable as both magnetic and turbulent forces are present during the gravitational collapse process.  In our simulation, we have identified twelve disk-like cores with axis ratio $>$ 2:1 and large enough for statistical analysis.  The sizes and mass of the twelve cores are in the range of young protostellar objects observed with flattened disk-like structure (Table 1).  All the disk-like cores are rotationally supported.  Magnetic pressure support is relatively unimportant for these cores, perhaps due to the large $M/M_{\rm cr}$ ratio that we chose as an initial condition.  The surface density of most of the disk-like cores has a power law distribution, as observed in real cores, and as predicted in isolated core collapse simulations.  Instead of being tidally truncated or disrupted, the core disks survive and flourish among strong interactions.

Most of the disk-like cores appear to be in Keplerian rotation from their PV diagram.  Observationally, many protostellar disks are found to be in Keplerian rotation \citep[e.g.][and references therein]{wis01}.  There are still some observed protostellar disks found to be in non-Keplerian rotation \citep[e.g.][]{mye00}.  \citet{wis01} points out that the non-Keplerian rotation curve reflects a mass distribution that is not yet dominated by a central condensation.  Also, a system could be in Keplerian rotation but a high optical depth would result in velocity maps that are dominated by the motion in the outer layers along the line of sight toward the center of the core.  For the twelve rotationally supported disk-like cores, a large amount of matter is still accreting perpendicular to the disks (Figure 21a); converted to astrophysical units this corresponds to substantial accretion rates (see Table 1).  When the mass of the disk keeps increasing, the gravitational stability of the rotational disk may be affected.  Unfortunately, because of the limited resolution, we cannot follow the core collapse long enough to see consequences beyond the first hint of spiral structure in the disk.

There is as yet no direct detection of disks around high-mass ($>$ 1 M$_{\sun}$) protostars.  \citet{fue01} suggests that for stellar mass $>$ 5 M$_{\sun}$, destruction mechanisms like radiation pressure on dust or ionization by stellar ultraviolet emission may be responsible for the rapid erosion of the outer disk.  However, these mechanisms would only operate at the later stages of protostellar evolution.  \citet{zha98} point out that the lack of direct detections of disks around high-mass protostars may simply be due to the greater distances to high mass star formation regions, as well as their clustering, which further complicates the kinematics.  Therefore, the existence of disks around high-mass protostars at the very early stage of core collapse remains uncertain.  It is important to note that in a number of objects studied at high resolution with radio and millimeter-wave interferometry, flattened structures and Keplerian motions argue strongly for the existence of rotationally supported disks \citep[e.g.][]{sar91,koe93,dut96}.  Improvements in the resolution of both observations and simulations on the early stage of core collapse in molecular clouds are vital to understand the details of star formation inside magnetized turbulent molecular clouds.

\acknowledgments
M.-M. M. L. acknowledges partial support by NASA Astrophysical Theory Program grant NAG5-10103 and NSF CAREER grant AST 99-85392.  F. Heitsch is supported by a Feodor-Lynen fellowship of the Alexander-von-Humboldt Foundation and by NSF grant AST-0098701.  We thank R. S. Klessen, P. Padoan, and E.G. Zweibel for useful discussions.  Computations were performed at the NCSA using the 256-processor SGI/Cray Origin 2000 through NRAC allocation QUS.

\clearpage

\begin{figure}
\centering
\includegraphics[scale=0.6]{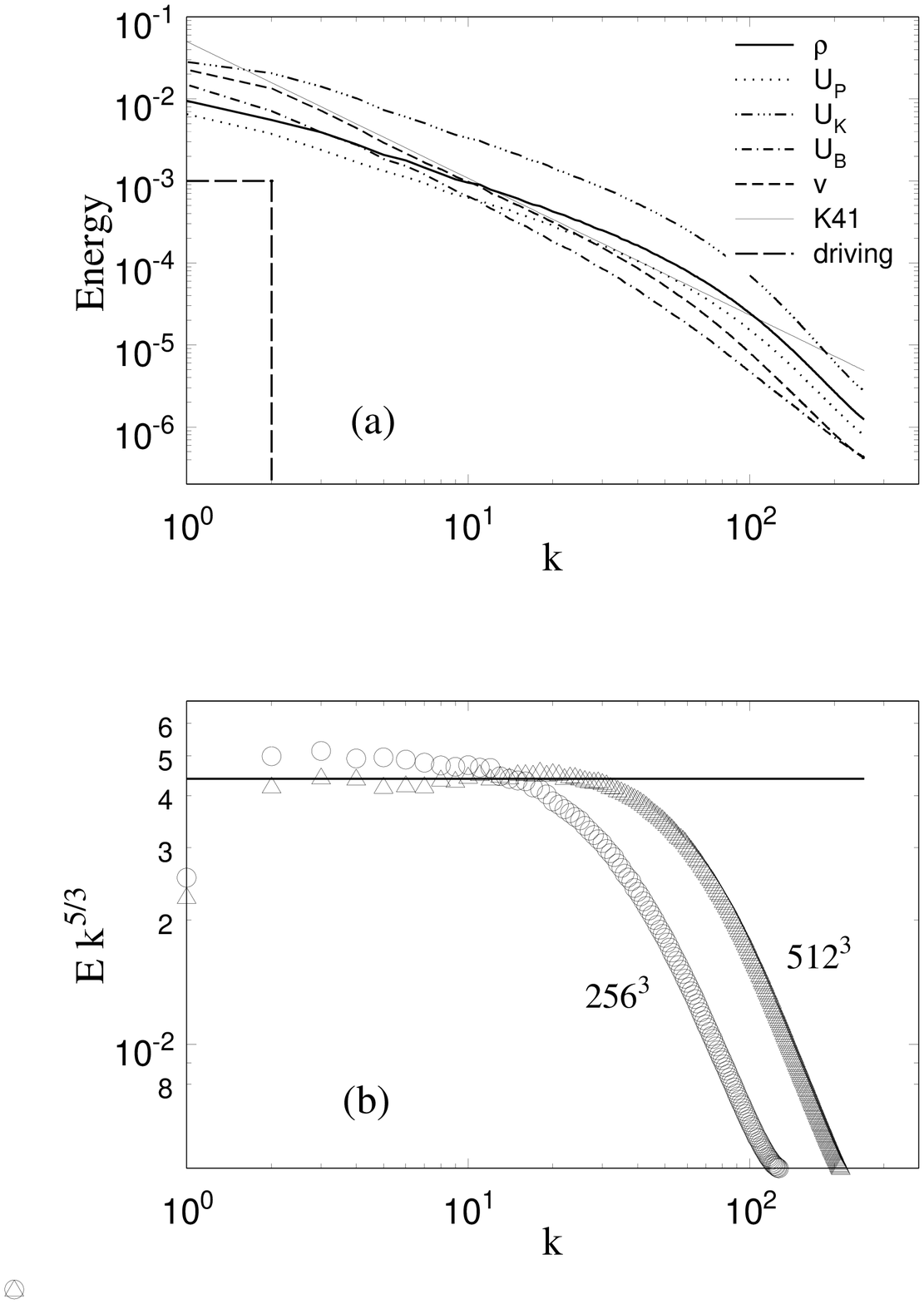}
\caption{(a) Power spectra of density $\rho$ (solid line), velocity $v$ (dashed line), potential energy $U_P$ (dotted line), kinetic energy $U_K$ (dash-dot-dot-dotted line), and magnetic energy $U_B$ (dash-dotted line) of the 512$^3$ simulation.  The thin straight line is the Kolmogorov power spectrum, which fits well to the inertial range of velocity and magnetic energy power spectra.  The long dashed line is the driving spectrum.  (b) Velocity power spectra of 256$^3$ (circle) and 512$^3$ (triangle) simulations of the turbulence, compensated by $k^{5/3}$, at $t$ = 0 when gravity is turned on. The horizontal straight line is the Kolmogorov power spectrum. \label{fig1}}
\end{figure}

\clearpage 

\begin{figure}
\includegraphics[scale=0.5,angle=-90]{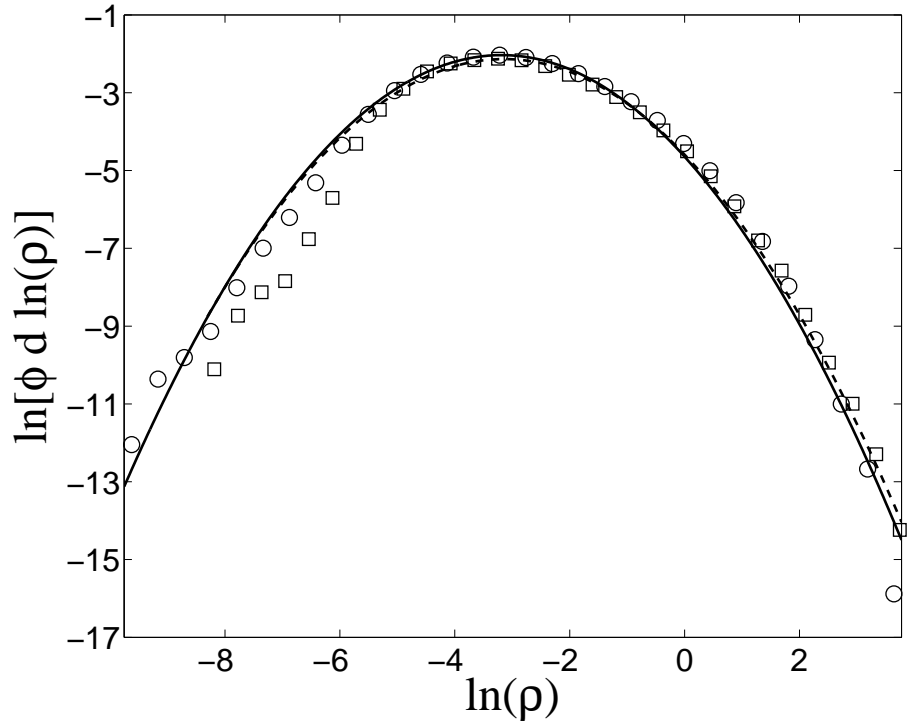}
\caption{Density PDF and log-normal fit in 256$^3$ (open square and dashed line) and 512$^3$ (open circle and solide line) simulations at $t = 0$. \label{fig2}}
\end{figure}

\clearpage 

\begin{figure}
\centering
\includegraphics[scale=0.8]{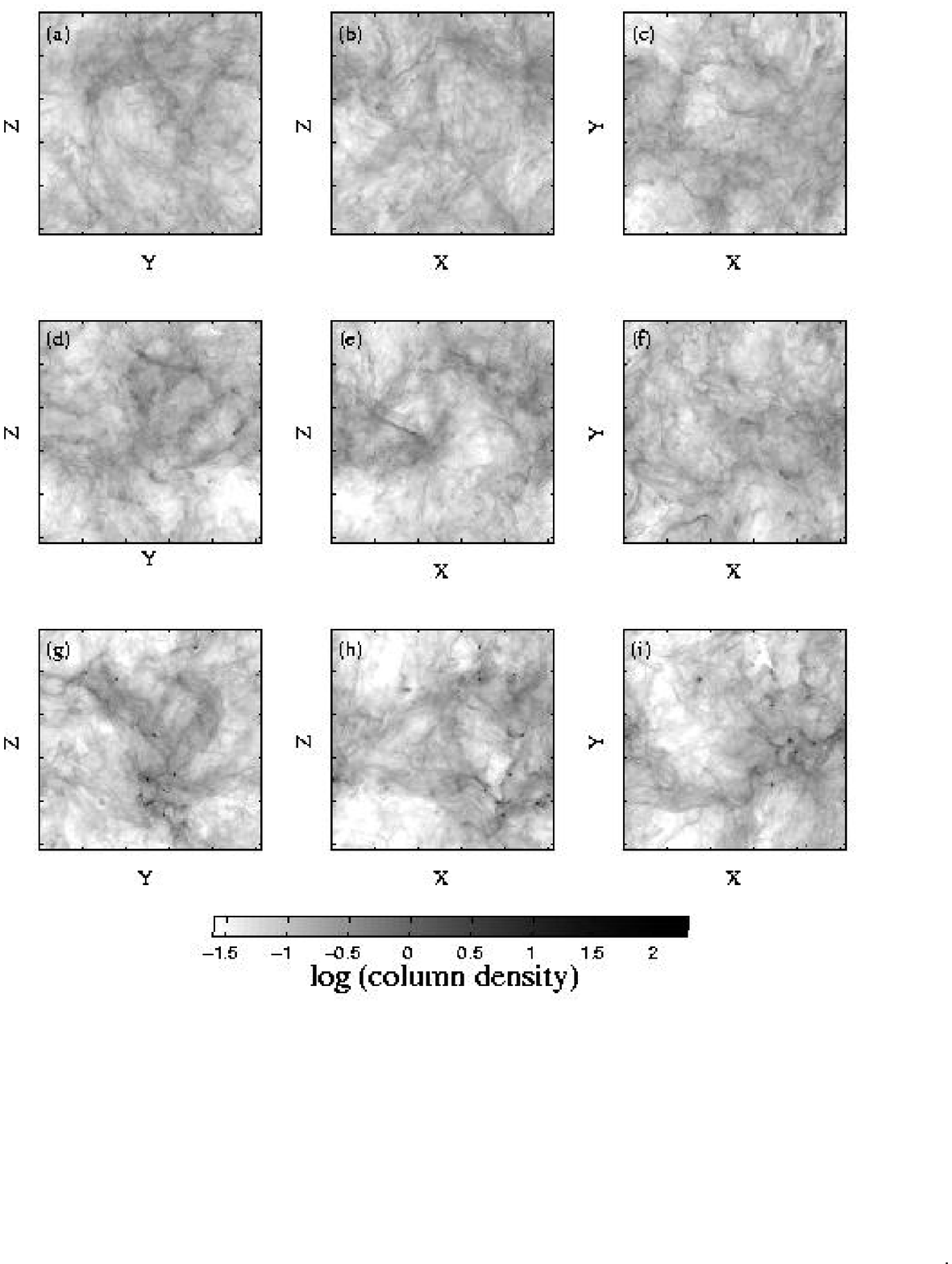}
\caption{Column density of the 512$^3$ simulation projected along the x-, y-, and z-directions.  All images are plotted using the same gray scale.  A log column density of -0.602, corresponds to gas at the mean density times the box length.  (a - c) At $t$ = 0, when gravitation is switched on, filamentary structure dominates the appearance.  (d - f) At $t = 0.65 \tau_{\rm ff}$, filaments break into pieces and higher density clumps slowly increase their mass by accreting material around them.  (g - i) At $t = 1.31 \tau_{\rm ff}$, many cores with high column-density contrast are formed. \label{fig3}}
\end{figure}

\clearpage 

\begin{figure}
\centering
\includegraphics[scale=0.5]{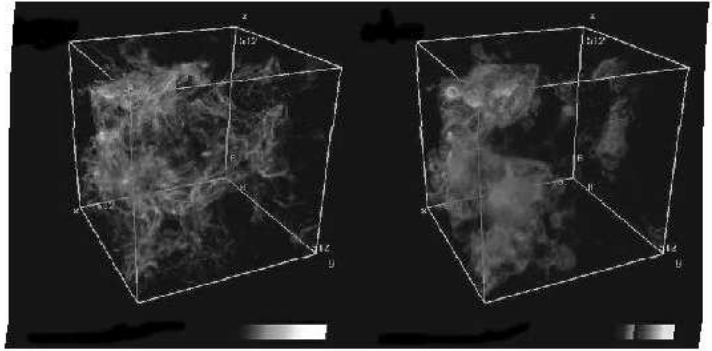}
\caption{3D volume rendering of the logarithm of density (left) and the magnetic pressure (right) of the turbulence box at $t = 1.31 \tau_{\rm ff}$.  The whole simulation is also available as an mpeg animation in system time unit. \label{fig4}}
\end{figure}

\clearpage 

\begin{figure}
\centering
\includegraphics[scale=0.5,angle=-90]{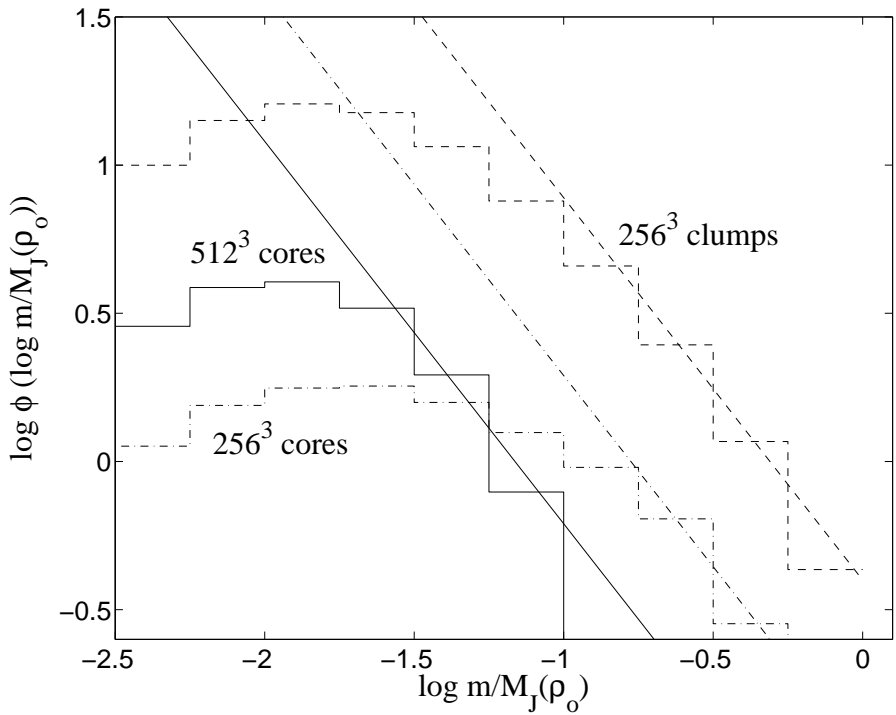}
\caption{The mass spectrum of cores (dot-dashed line) and clumps (dashed line), in the 256$^3$ simulation and the mass spectrum of cores (solid line) in the 512$^3$ simulation at $t$ = 0.  The spectra are normalized by the Jeans mass, $M_{\rm J}(\rho)$.  The straight lines are the power law with index -1.29 predicted by \citet{pad02} for our velocity power spectrum.  High-mass wings of all 3 spectra are consistent with the predicted power law.  \label{fig5}}
\end{figure}

\clearpage 

\begin{figure}
\centering
\includegraphics[scale=0.5,angle=-90]{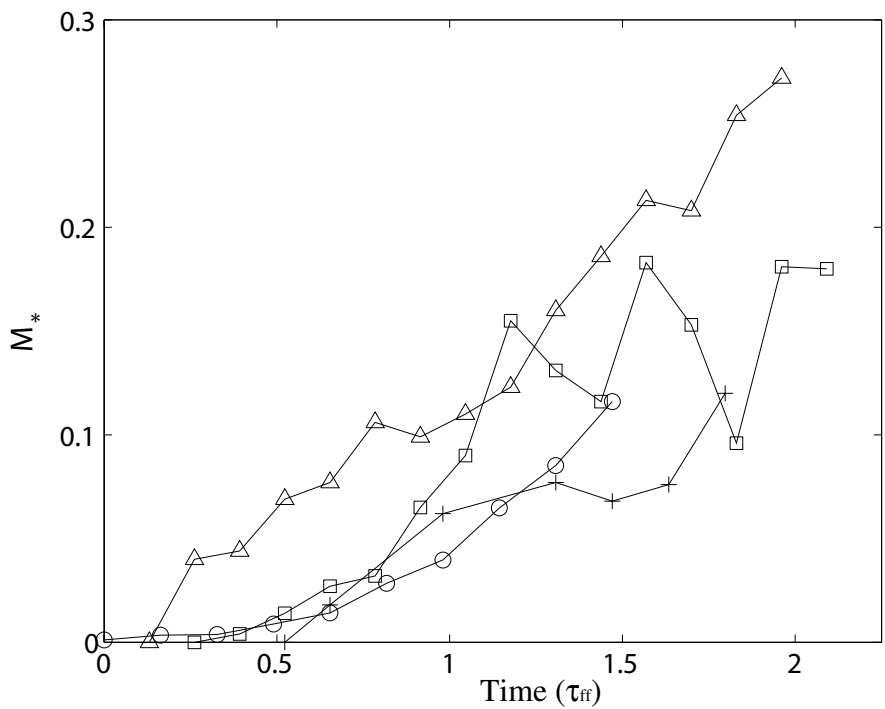}
\caption{Comparison of the mass in gravitationally bound cores for runs driven at $k$ = 1 - 2 with varying resolution 64$^3$ (triangle), 128$^3$ (square), 256$^3$ (plus), and 512$^3$ (circle).  $M_*$ denotes the sum of masses found in all cores, in units of box mass, determined by the modified CLUMPFIND \citep[][; see also Paper I]{wil94}.  Collapse rates vary but collapse occurs in all cases. \label{fig6}}
\end{figure}

\clearpage 

\begin{figure}
\centering
\includegraphics[scale=0.5,angle=-90]{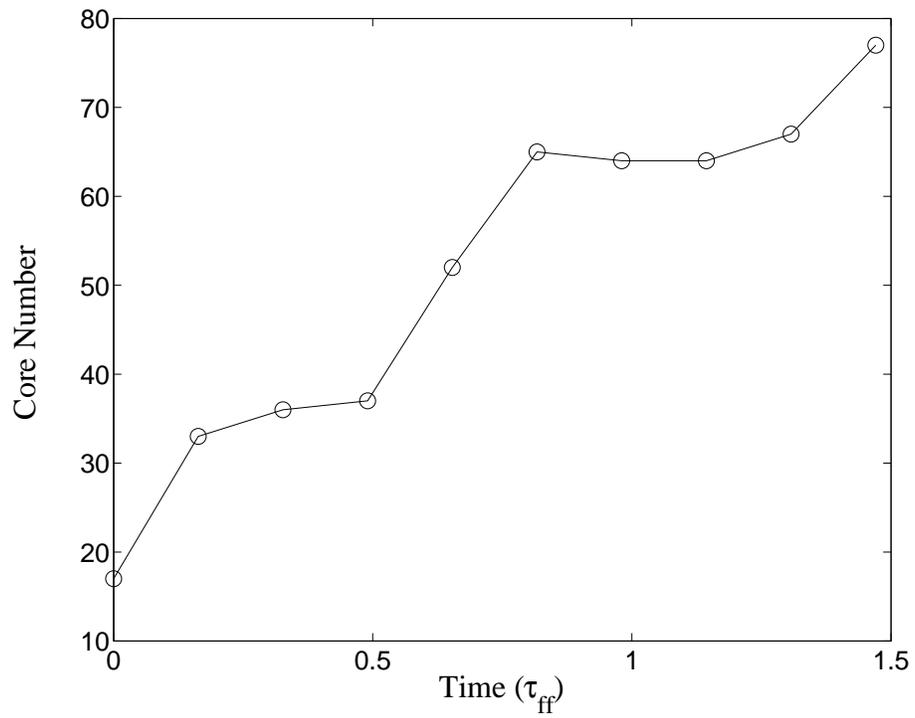}
\caption{Core number counted from the simulation using the modified CLUMPFIND.  The decrease in core number is caused by the destruction of cores by supersonic turbulence or merging of cores. \label{fig7}}
\end{figure}
\clearpage 

\clearpage

\begin{figure}
\centering
\includegraphics[scale=0.7,angle=-90]{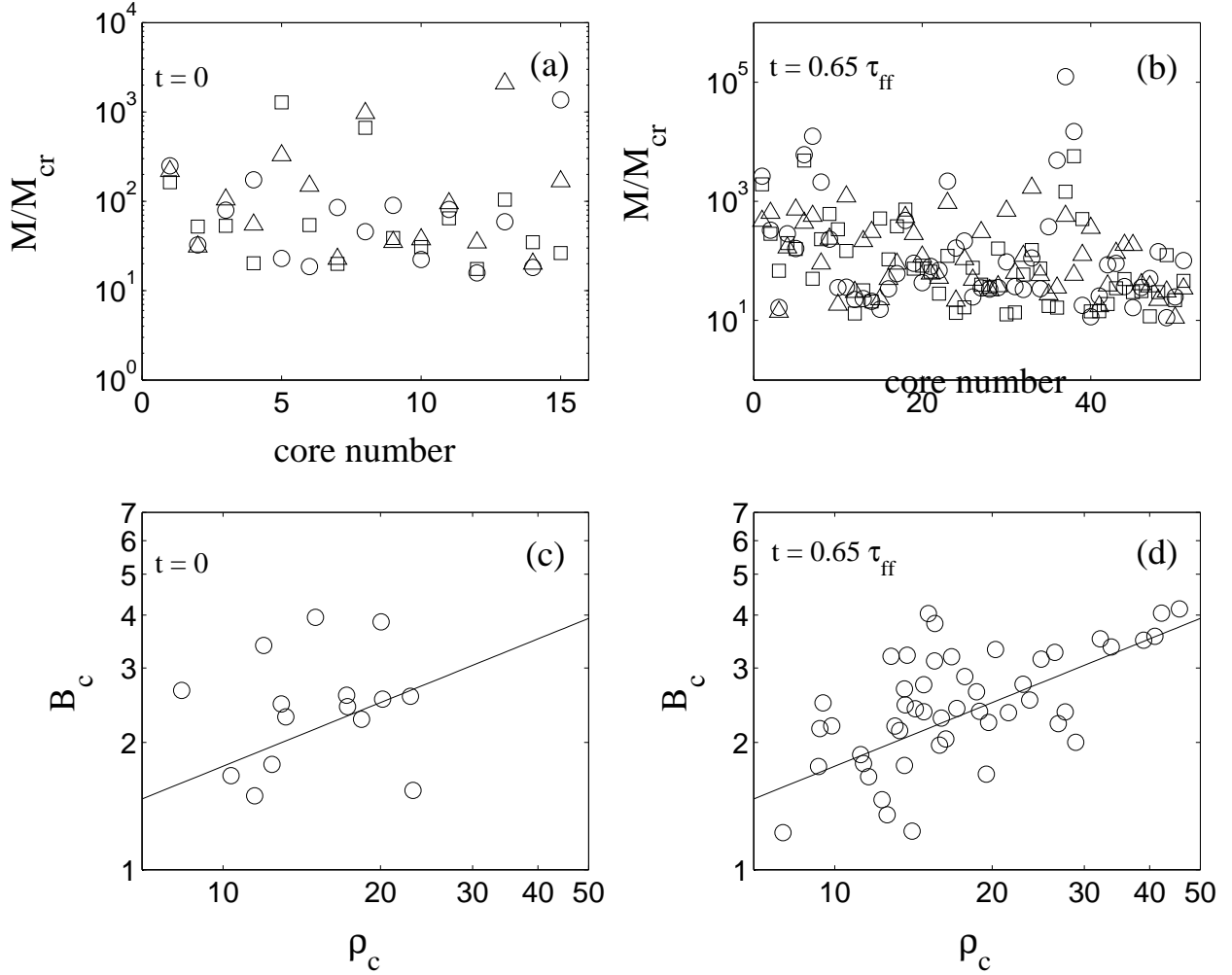}
\caption{Evolution of magnetic field and density of cores.  Ratio of core mass to critical mass for magnetostatic support $M/M_{\rm cr}$ is shown for cores at (a) $t = 0$, and (b) $t = 0.65 \tau_{\rm ff}$ in x- (circle), y- (triangle), and z-direction (square).  Cores are magnetically supercritical.  Central magnetic field strength $B_c$ vs. central density of cores $\rho_c$ is shown at (c) $t$ = 0, and (d) $t = 0.65 \tau_{\rm ff}$.  The straight line is a power law of index 0.5. \label{fig8}}
\end{figure}

\clearpage 

\begin{figure}
\centering
\includegraphics[scale=0.7,angle=-90]{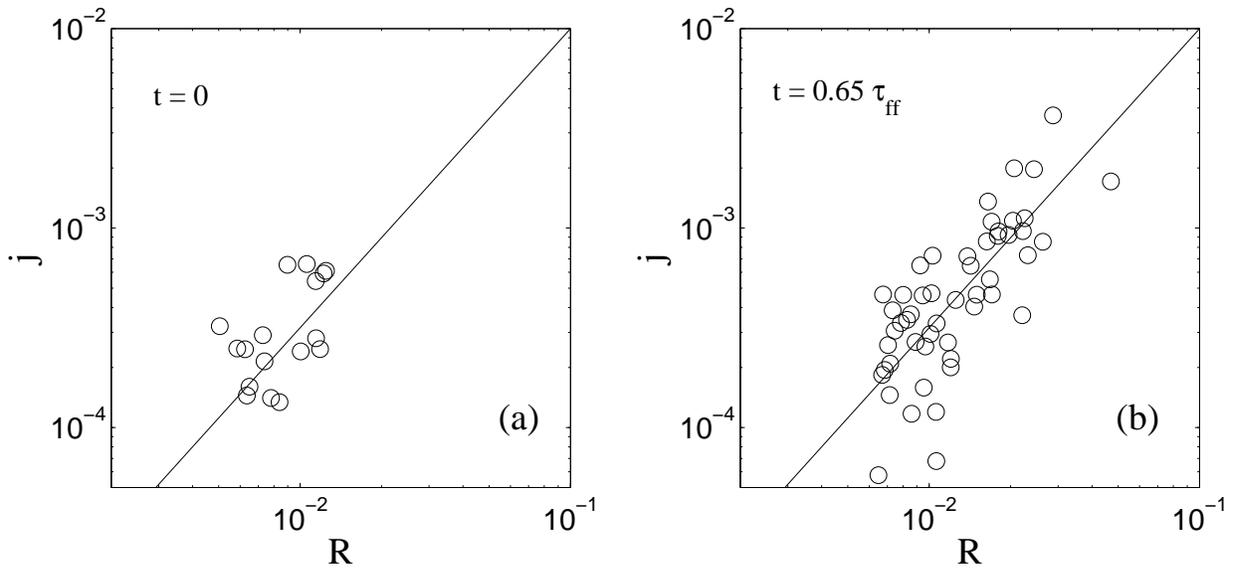}
\caption{Evolution of specific angular momentum j vs. radius R of cores at (a) $t$ = 0, and (b) $t = 0.65 \tau_{\rm ff}$.  The straight line is a power law of index 3/2.  The data distribution matches well with observations of dark cloud cores \citep[e.g.][]{goo93}. \label{fig9}}
\end{figure}

\clearpage 

\begin{figure}
\centering
\includegraphics[scale=0.7,angle=-90]{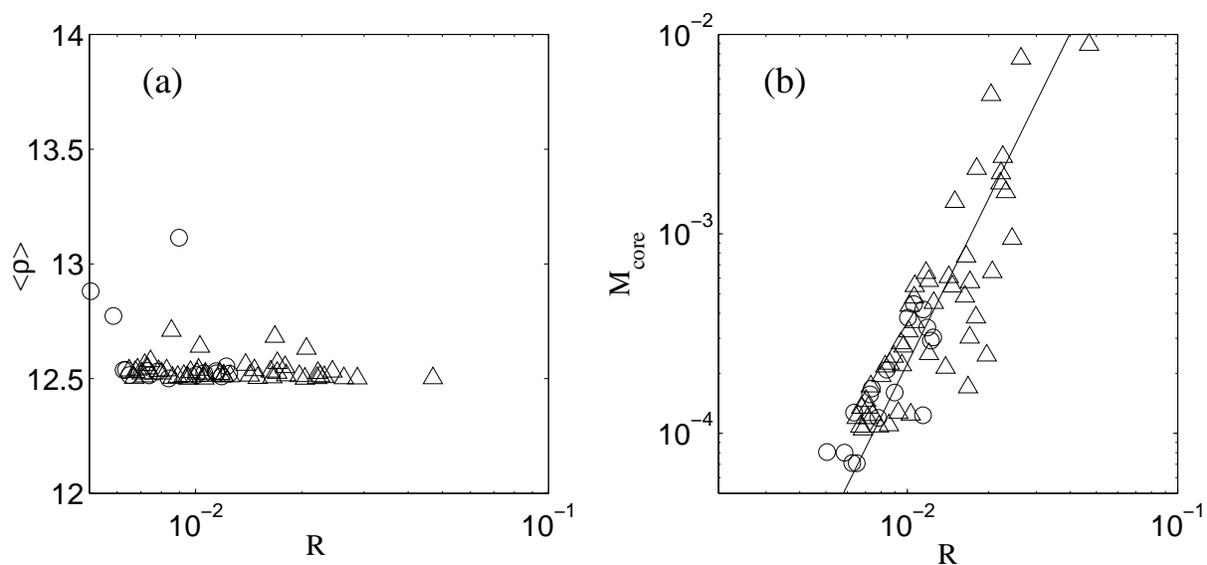}
\caption{(a) Average density $<\rho>$ vs. core radius $R$ at $t$ = 0 (open circles) and $t$ = 0.65 $\tau_{\rm ff}$ (open triangle).  Basically, core radius is unrelated to core average density \citep{bal02}.  (b) Core radius $R$ vs. core mass $M_{core}$ at $t$ = 0 (open circles) and $t = 0.65 \tau_{\rm ff}$ (open triangle).  The straight line is a power law of index 2.75.  All quantities are in scaled system units. \label{fig10}}
\end{figure}

\clearpage 

\begin{figure}
\centering
\includegraphics[scale=0.8]{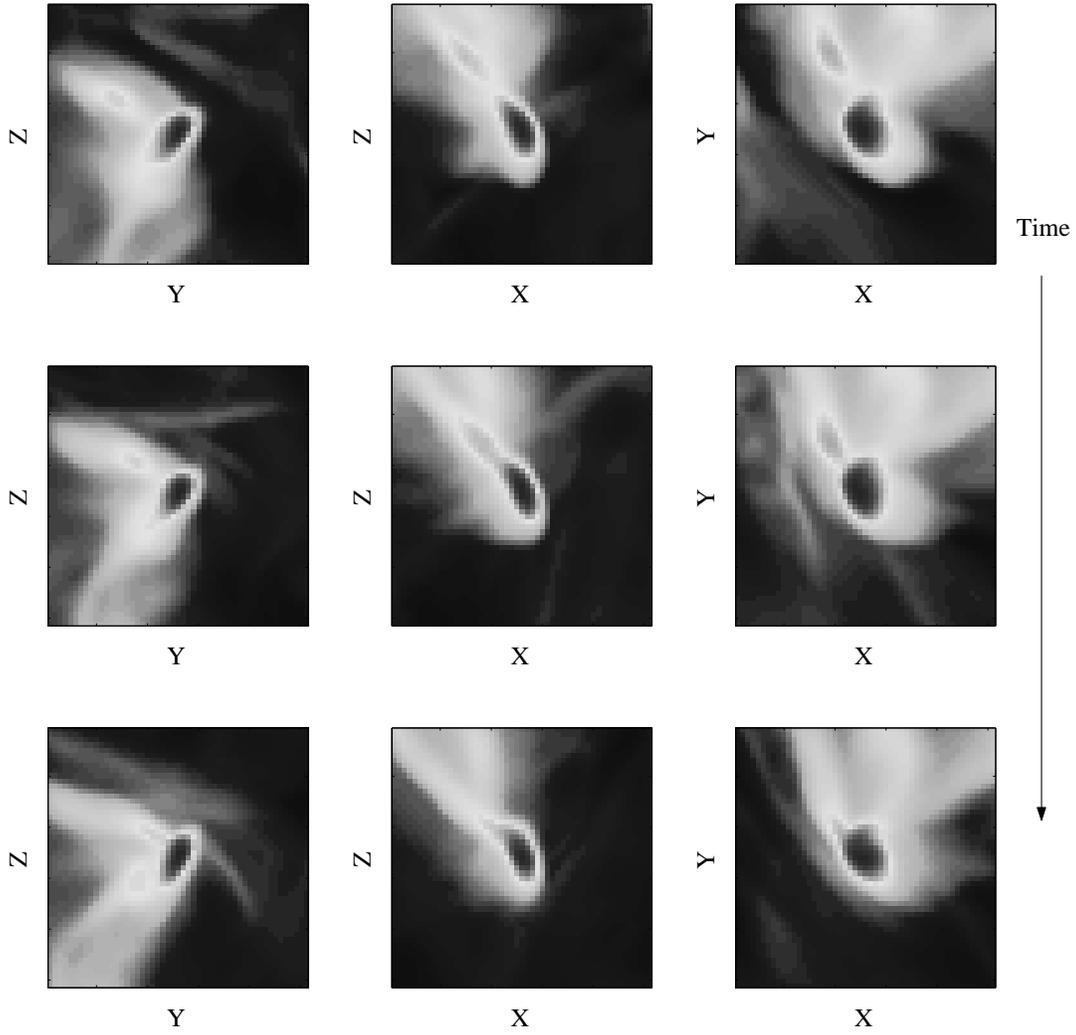}
\caption{Different views of a core accretion event along the x-, y-, and z-directions.  The time interval between two rows of the figures is 0.05 $\tau_{\rm ff}$.  Core merging and accretion of large material clump are an important process in forming disk-like cores. \label{fig11}}
\end{figure}

\clearpage 

\begin{figure}
\centering
\includegraphics[scale=0.8]{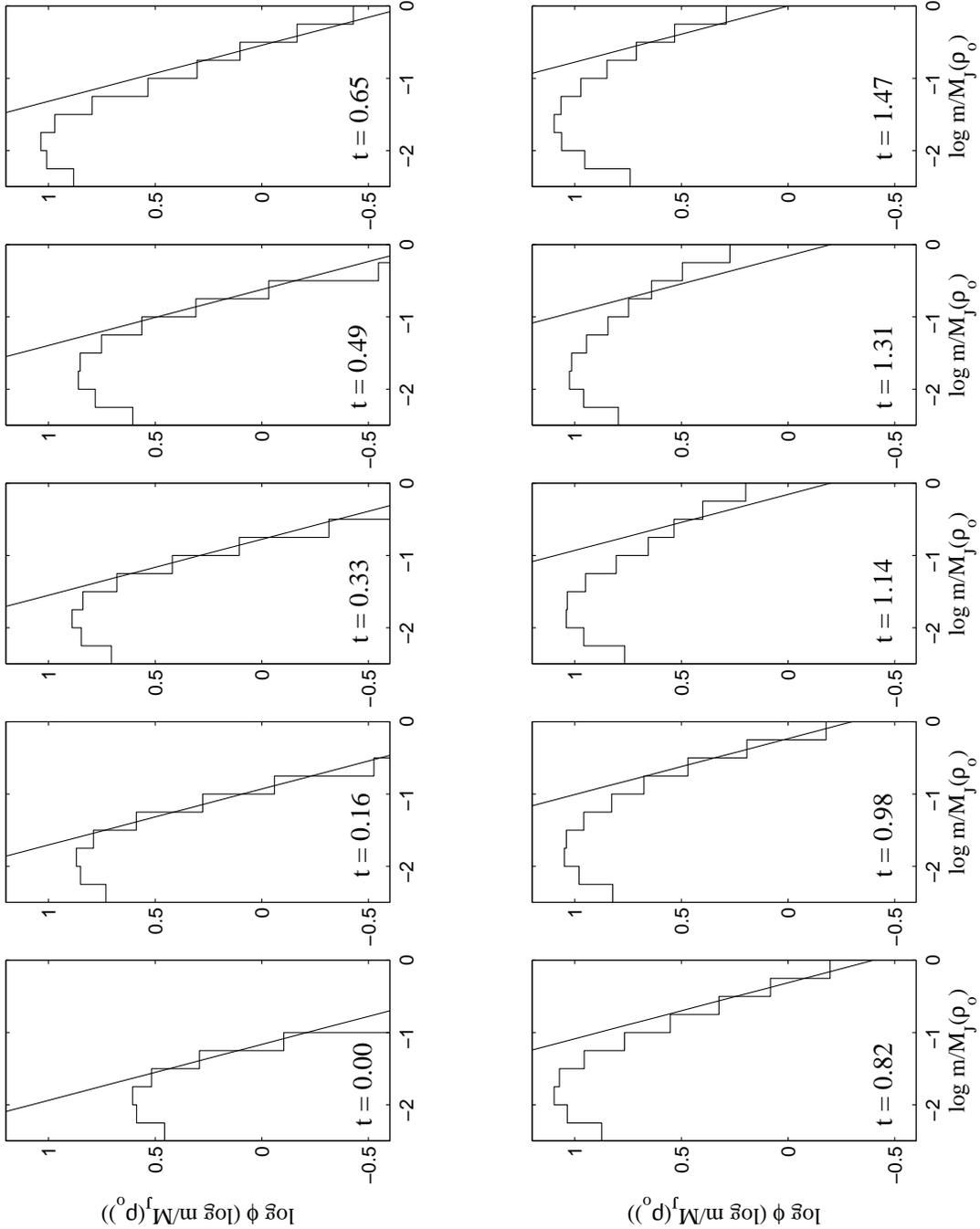}
\caption{Evolution of the core mass spectrum, normalized by the Jeans mass $M_{\rm J}(\rho_o)$, in the 512$^3$ simulation.  The straight line is the power law with index -1.29 predicted by \citet{pad02} for $\beta$ = -5/3 (see text).  The core mass spectra generally show a "universal" IMF appearance \citep{kro02} in the simulation with a clear turn over at the low mass region.  The slope of the high-mass wing of the spectra matches well with the turbulent fragmentation prediction and remain about the same until $t = 0.49 \tau_{\rm ff}$.  Later the slope becomes shallower as the result of core coalescence.  \label{fig12}}
\end{figure}

\clearpage 

\begin{figure}
\centering
\includegraphics[scale=0.5,angle=-90]{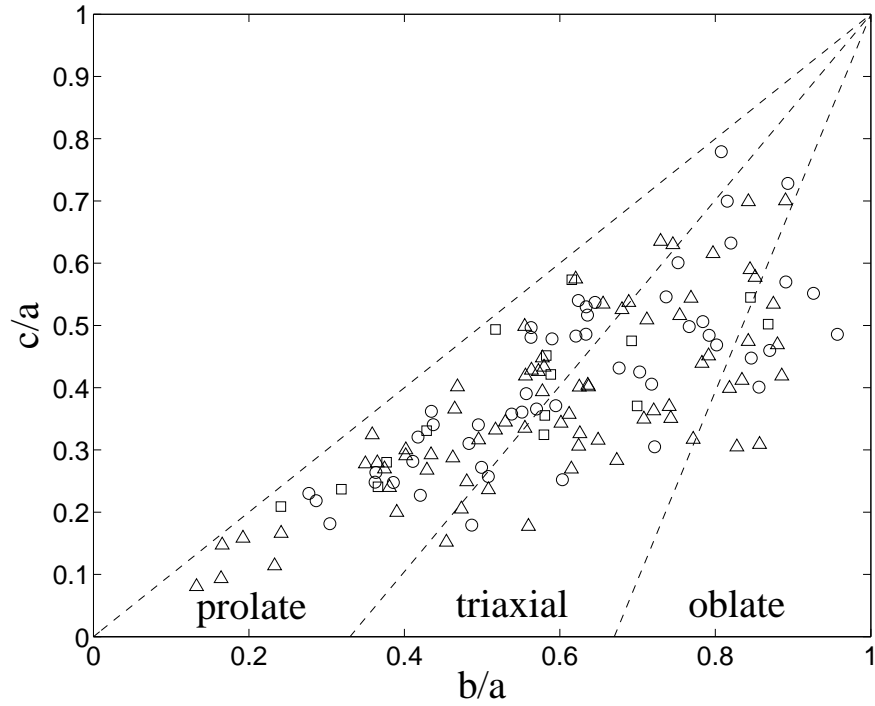}
\caption{The distribution of core axis ratios at $t$ = 0 (square), 0.65 (circle), and 1.31 $\tau_{\rm ff}$ (triangle).  Cores near the diagonal are prolate and the cores near $b/a$ = 1 are oblate.  The region is divided into 3 parts and the cores inside the middle region are classified as triaxial \citep{gam03}.  The majority of cores are prolate or triaxial in shape.  The principal axes of cores are defined such that $a > b > c$. \label{fig13}}
\end{figure}

\clearpage 

\begin{figure}
\centering
\includegraphics[scale=0.5,angle=-90]{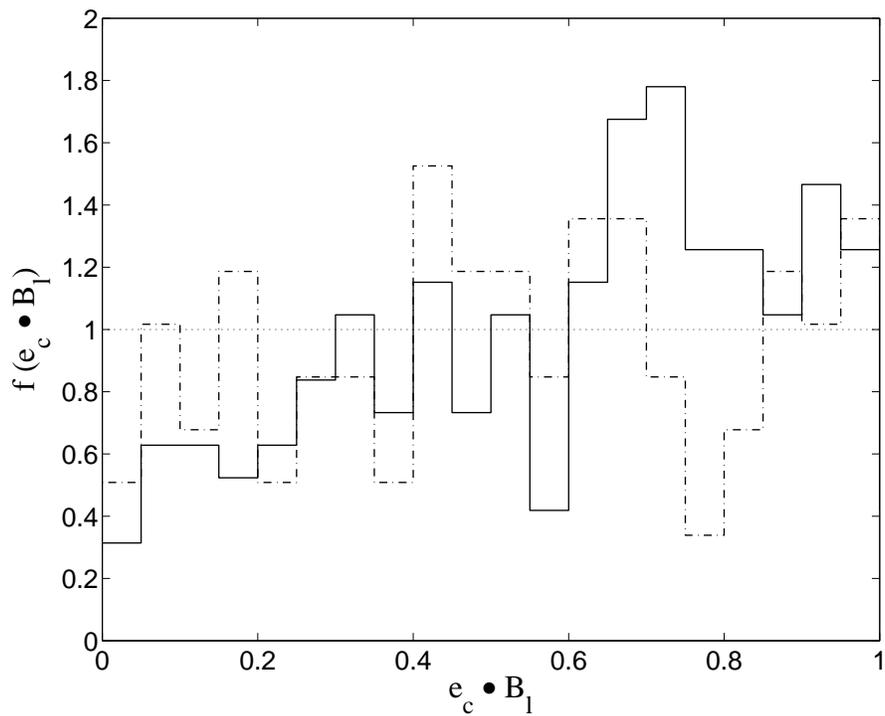}
\caption{The correlation of angles between the shortest body axis (\^{e}$_{\rm c}$)and the density weighted mean magnetic field ($B_l$) in the cores at $t = 0.65 \tau_{\rm ff}$ (dot-dashed line) and 1.31 $\tau_{\rm ff}$ (solid line).  We observe stronger correlation of the angles in this simulation than that in the simulation by \citep{gam03} with similar initial weak field. \label{fig14}}
\end{figure}

\clearpage 

\begin{figure}
\centering
\includegraphics[scale=1.8]{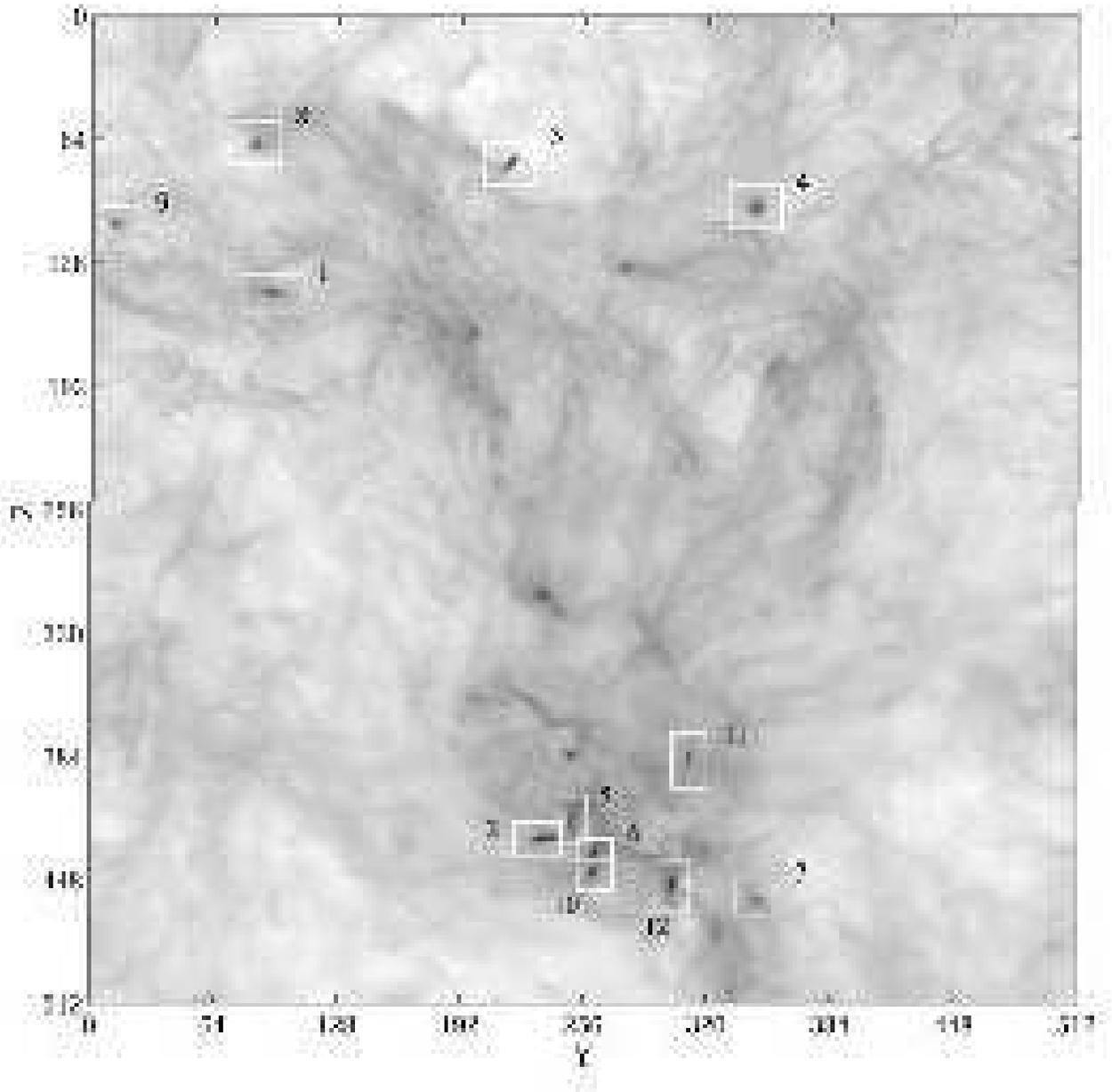}
\caption{Column density map highlighting the location of 12 disk-like cores at $t = 1.31 \tau_{\rm ff}$.  The properties of core 1 is analyzed in detail in \S5.3. \label{fig15}}
\end{figure}

\clearpage 

\begin{figure}
\centering
\includegraphics[scale=0.7]{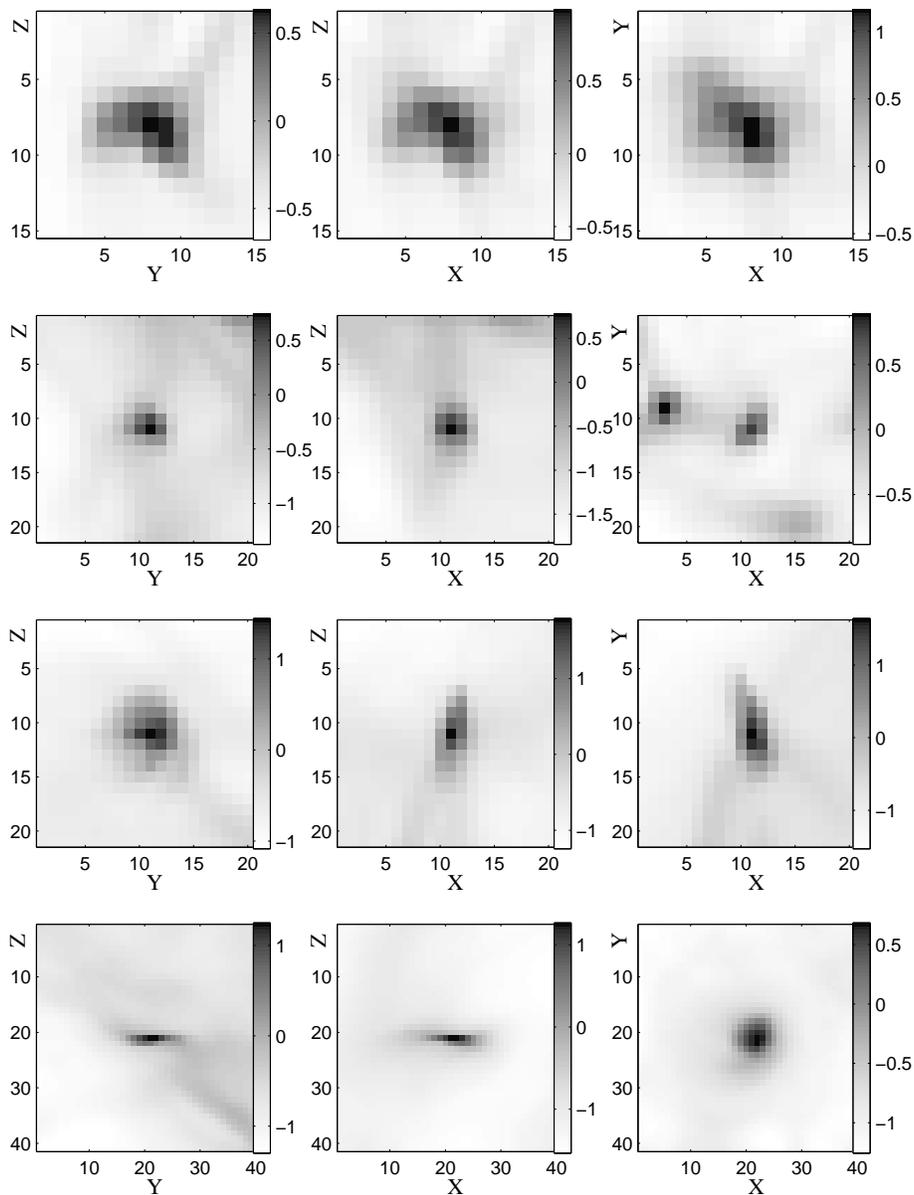}
\caption{The effect of numerical resolution on core structure when total 10\% of material is trapped inside cores: 64$^3$ (1st row), 128$^3$ (2nd row), 256$^3$ (3rd row), and 512$^3$ (4th row).  Grey scale images of column density, projected along x-, y-, and z-axes and the axes denote cell number.  Disk-like core structure barely seen in 256$^3$ resolution and becomes clear in the 512$^3$ simulation. \label{fig16}}
\end{figure}

\clearpage 

\begin{figure}
\centering
\includegraphics[scale=0.5]{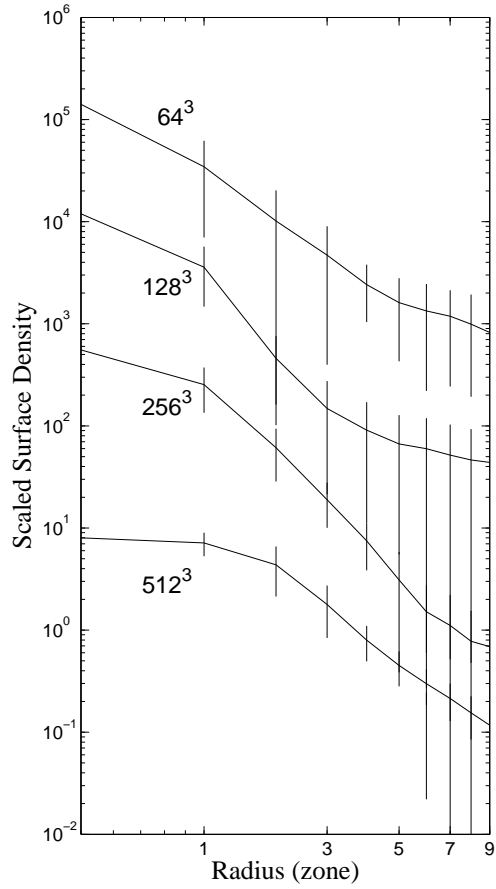}
\caption{Surface density of the cores in four simulations with resolutions 64$^3$, 128$^3$, 256$^3$, and 512$^3$, as shown in Figure 16.  The surface densities of the cores in 64$^3$, 128$^3$, and 256$^3$ resolution simulations are scaled by 10$^4$, 10$^3$, and 10, respectively for comparison.  With higher resolution, the core surface density profile converges to a uniform density central region and a power law disk. \label{fig17}}
\end{figure}

\clearpage 

\begin{figure}
\centering
\includegraphics[scale=0.6,angle=-90]{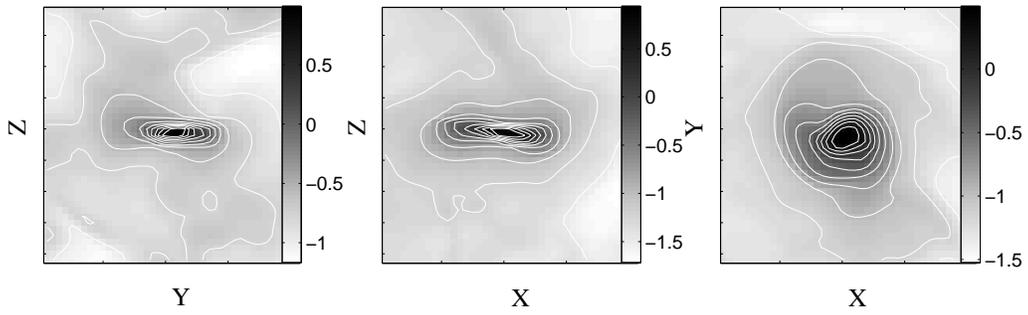}
\caption{Logarithmic column density images of Core 1 in 3 directions in gray scale and contours.  The contour interval is 0.265.  A spiral structure of the accretion disk is barely visible in the z-direction. \label{fig18}}
\end{figure}

\clearpage 

\begin{figure}
\centering
\includegraphics[scale=0.6,angle=-90]{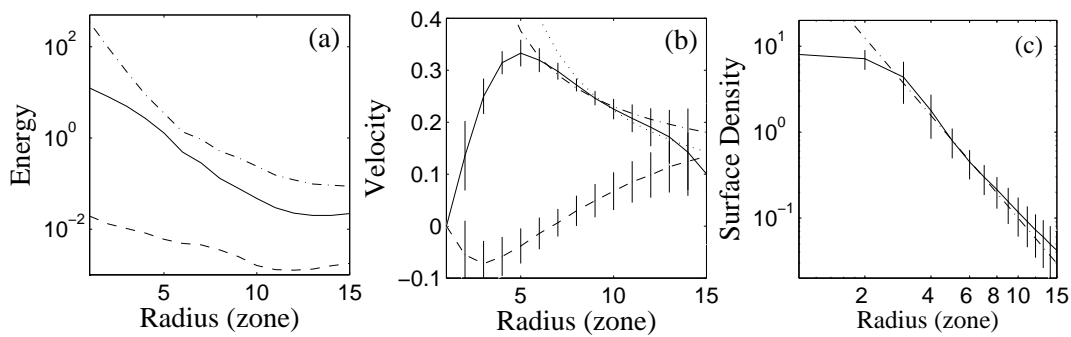}
\caption{Radial profiles of spherically averaged properties of core 1 at $t = 1.31 \tau_{\rm ff}$.  (a) Kinetic (solid line), magnetic (dashed line), and potential (dot-dashed line) energy distribution along the radius. (b) Rotational (solid line) and radial (dashed line) velocity profiles.  The theoretical rotational curves for a Keplerian rotation (dot-dashed line) and angular momentum conservation rotation (dotted line) are also shown.  Keplerian rotation seems to fit better to the data. (d) Surface density of the core shows a power law outer profile.  The slope of the power law fit (dot-dashed line) is -3. \label{fig19}}
\end{figure}

\clearpage 

\begin{figure}
\centering
\includegraphics[scale=0.7,angle=-90]{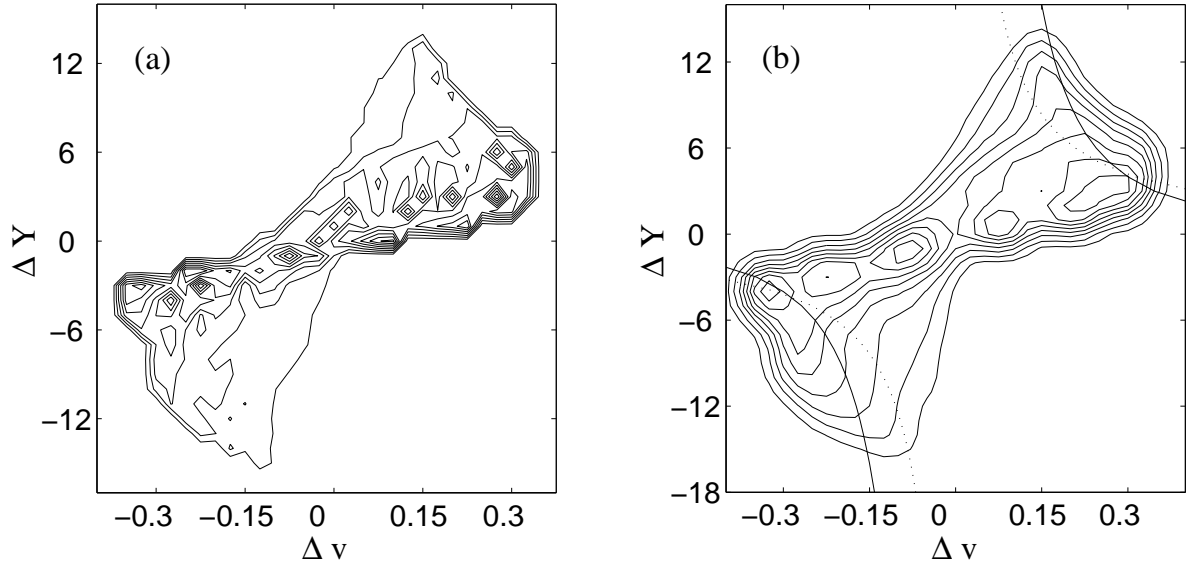}
\caption{Position-velocity diagrams of the Core 1.  (a) PV diagram before convolution.  The contours are the logarithm of column density.  (b) PV diagram after convolving (a) with a circular Gaussian beam with a FWHM of 3 zones.  The solid curve indicates the Keplerian rotation and dotted curve indicates the angular momentum conservation rotation. \label{fig20}}
\end{figure}

\clearpage 

\begin{figure}
\centering
\includegraphics[scale=0.7,angle=-90]{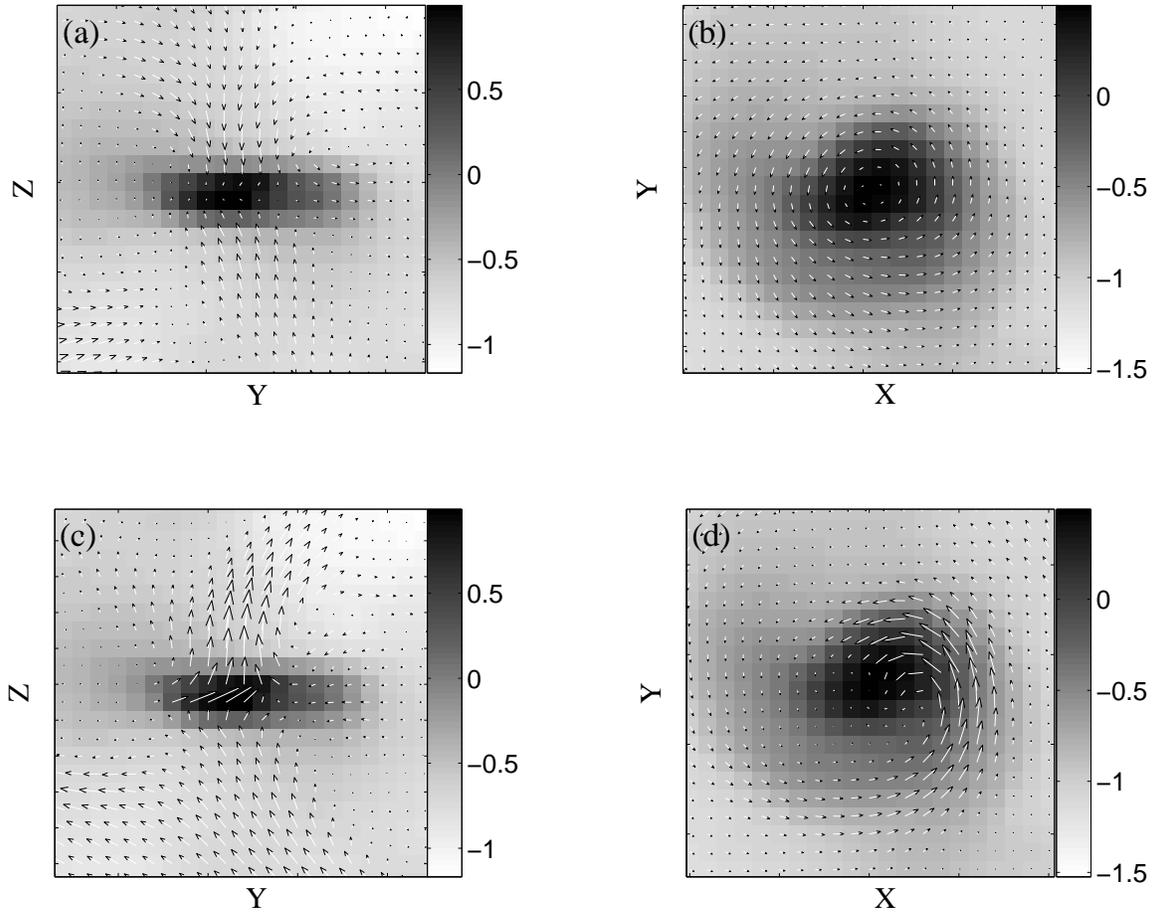}
\caption{Vector plots of velocity and magnetic field in Core 1.  (a) Side view along the x-direction shows material accretion along the polar direction.  (b) Plan view along z-direction shows disk rotation.  (c) Side view along the x-direction shows a distorted bipolar structure of the magnetic field.  (d) Plan view along z-direction shows dragging of magnetic field by the core rotation.  Surface density images are shown in gray scale. \label{fig21}}
\end{figure}

\clearpage 

\begin{figure}
\centering
\includegraphics[scale=0.3]{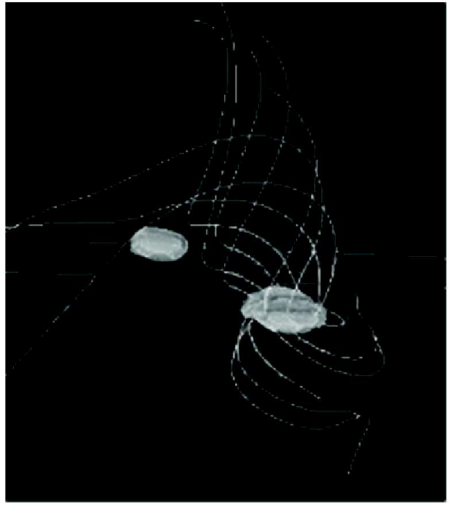}
\caption{The 3D view of the helical magnetic field through core 6 caused by core rotation.  The solid rendered objects are density isosurfaces. \label{fig22}}
\end{figure}

\clearpage 

\begin{deluxetable}{cccccccc}
\tabletypesize{\scriptsize}
\tablecaption{Physical properties of 12 disk-like cores. \label{tbl-1}}
\tablewidth{0pt}
\tablehead{
\colhead{Core} &\colhead{Category} &\colhead{Mass\tablenotemark{a}} &\colhead{Diameter\tablenotemark{a}} &\colhead{Accretion Rate\tablenotemark{a}} &\colhead{Mass\tablenotemark{b}} &\colhead{Diameter\tablenotemark{b}} &\colhead{Accretion Rate\tablenotemark{b}} \\
  &  &  & & & \colhead{(M$_\sun$)} & \colhead{(AU)} & \colhead{(M$_\sun$ yr$^{-1}$)} 

}
\startdata
1  &1 &0.0035 &0.0977 &5.25$\times$10$^2$ &0.457 &5642 &9.72$\times$10$^{-6}$ \\
2  &2 &0.0114 &0.0586 &2.18$\times$10$^3$ &1.489 &3384 &4.04$\times$10$^{-5}$ \\
3  &1 &0.0028 &0.0547 &3.94$\times$10$^2$ &0.366 &3159 &7.30$\times$10$^{-6}$ \\
4  &1 &0.0033 &0.0508 &2.07$\times$10$^3$ &0.431 &2934 &3.84$\times$10$^{-5}$ \\
5  &2 &0.0192 &0.0625 &3.91$\times$10$^3$ &2.508 &3610 &7.25$\times$10$^{-5}$ \\
6  &2 &0.0121 &0.0547 &2.73$\times$10$^3$ &1.580 &3159 &5.06$\times$10$^{-5}$ \\
7  &2 &0.0015 &0.0430 &1.26$\times$10$^3$ &0.196 &2483 &2.34$\times$10$^{-5}$ \\
8  &1 &0.0017 &0.0586 &1.53$\times$10$^3$ &0.222 &3384 &2.83$\times$10$^{-5}$ \\
9  &1 &0.0011 &0.0313 &1.05$\times$10$^3$ &0.144 &1808 &1.95$\times$10$^{-5}$ \\
10 &2 &0.0064 &0.0586 &1.90$\times$10$^3$ &0.836 &3384 &3.53$\times$10$^{-5}$ \\
11 &2 &0.0048 &0.0781 &2.22$\times$10$^3$ &0.627 &4511 &4.12$\times$10$^{-5}$ \\
12 &2 &0.0138 &0.0195 &5.59$\times$10$^3$ &1.802 &1126 &1.04$\times$10$^{-4}$ \\
\enddata

%% Text for table notes should follow after the \enddata but before
%% the \end{deluxetable}. Make sure there is at least one \tablenotemark
%% in the table for each \tablenotetext.

\tablenotetext{a} {Units in simulation scaling}
\tablenotetext{b} {Units in high density scaling}

\tablecomments{Cores in category 1 belong to a group with density, magnetic pressure, and accretion rate generally higher than those of category 2 cores found in open space.}

\end{deluxetable}


\begin{thebibliography}{}

\bibitem[Avillez(2000)]{avi00} de Avillez, M. A. 2000, \mnras, 315, 479
\bibitem[Balbus \& Hawley(1998)]{bal98} Balbus, S. A.  \& Hawley, J. F. 1998, Rev. Mod. Phys., 70, 1
\bibitem[Ballesteros-Paredes et al.(1999)]{bal99} Ballesteros-Paredes, J., Hartmann, L., \& V\'{a}zquez-Semadeni, E. 1999, \apj, 527, 285
\bibitem[Ballesteros-Paredes \& Mac Low(2002)]{bal02} Ballesteros-Paredes, J., Mac Low, M.-M. 2002, \apj, 570, 734
\bibitem[Barranco \& Goodman(1998)]{bar98} Barranco, J. A. \& Goodman, A. A. 1998, \apj, 504, 207
\bibitem[Basu(1997)]{bas97} Basu, S. 1997, \apj, 485, 240
\bibitem[Basu(2000)]{bas00} Basu, S. 2000, \apj, 540, L103
\bibitem[Basu \& Mouschovias(1994)]{bas94} Basu, S. \& Mouschovias, T. Ch. 1994, \apj, 432, 720
\bibitem[Basu \& Mouschovias(1995)]{bas95} Basu, S. \& Mouschovias, T. Ch. 1995, \apj, 453, 271
\bibitem[Bate, Bonnell, \& Bromm(2003)]{bat03} Bate, M. R., Bonnell, I. A., \& Bromm, V. 2003, \mnras, 339, 577
\bibitem[Bertoldi \& McKee(1992)]{ber92} Bertoldi, F. \& McKee, C. F. 1992, \apj, 395, 140
\bibitem[Bertoldi \& McKee(1997)]{ber97} Bertoldi, F. \& McKee, C. F. 1997, RevMexAA Conf. Ser., 6, 195
\bibitem[Blitz(1993)]{bli93} Blitz, L. 1993, in Protostars \& Planets III, eds. E. H. Levy \& J. I. Lunine (The University of Arizona Press, Tucson), 125
\bibitem[Boldyrev, Nordlund, \& Padoan(2002)]{bol02} Boldyrev, S., Nordlund, ${\rm \AA}$, \& Padoan, P. 2002, \apj, 573, 678
\bibitem[Bonnell et al.(1997)]{bon97} Bonnell, I. A., Bate, M. R., Clarke, C. J., \& Pringle, J. E. 1997, \mnras, 285, 201
\bibitem[Bonnell et al.(2001a)]{bon01a} Bonnell, I. A., Bate, M. R., Clarke, C. J., \& Pringle, J. E. 2001a, \mnras, 323, 785
\bibitem[Bonnell et al.(2001b)]{bon01b} Bonnell, I. A., Clarke, C. J., Bate, M. R., \& Pringle, J. E. 2001b, \mnras, 324, 573
\bibitem[Bonnell \& David(1998)]{bon98} Bonnell, I. A. \& David M. B. 1998, \mnras, 295, 691
\bibitem[Bourke et al.(2001)]{bou01} Bourke, T. L., Myers, P. C., Robinson, G., \& Hyland, A. R. 2001, \apj, 554, 916
\bibitem[Burgers(1974)]{bur74} Burgers, J. M. 1974, The Nonlinear Diffusion Equation (Dordrecht: Reidel)
\bibitem[Burkert \& Bodenheimer(2000)]{bur00} Burkert, A. \& Bodenheimer, P. 2000, \apj, 543, 822
\bibitem[Cho, Lazarian, \& Vishniac(2003)]{cho03} Cho, J., Lazarian, A., \& Vishniac, E. T. 2003, in Turbulence and Magnetic Fields in Astrophysics, eds. Falgarone, E. \& Passot, T. (Springer-Verlag: New York), 56
\bibitem[Clarke \& Pringle(1991)]{cla91} Clarke, C. J. \& Pringle, J.E. 1991, \mnras 249, 584
\bibitem[Crutcher(1999)]{cru99} Crutcher, R. M. 1999, \apj, 520, 706
\bibitem[Dewar(1970)]{dew70} Dewar, R. L. 1970, Phys. Fluids, 13, 2710
\bibitem[Dutrey(1996)]{dut96} Dutrey, A. 1996, in Science with Large Millimeter Arrays, ed. P. A. Shaver (New York: Springer), 235
\bibitem[Elmegreen(2000)]{elm00} Elmegreen, B. G. 2000, \apj, 530, 277
\bibitem[Fiege \& Pudritz(1999)]{fie99} Fiege, J. D., \& Pudritz, R. E. 1999, in ASP Conf. Proc. 168, New Perspectives on the Interstellar Medium, eds. A. R. Taylor, T. L. Landecker, \& G. Joncas (San Francisco: ASP), 248
\bibitem[Fleck (1981)]{fle81} Fleck, R. C., Jr. 1981, \apj, 246, L151
\bibitem[Frigo \& Johnson(1998)]{fri98} Frigo, M. \& Johnson, S. G. 1998, International Conference on Acoustics Speech and Signal Processing, 3, 1381
\bibitem[Fuente et al.(2001)]{fue01} Fuente, A., Neri, R., Martín-Pintado, J., Bachiller, R., Rodríguez-Franco, A., \& Palla, F. 2001, \aap, 366, 873
\bibitem[Gammie et al.(2003)]{gam03} Gammie, C. F., Lin, Y. T., Stone, J. M., \& Ostriker, E. C. 2003, \apj, 592, 203
\bibitem[Goldreich \& Sridhar(1995)]{gol95} Goldreich, P. \& Sridhar, S. 1995, \apj, 438, 763
\bibitem[Goodman et al.(1993)]{goo93} Goodman, A. A., Benson, P. J., Fuller, G. A., \& Myers, P. C. 1993, \apj, 406, 528
\bibitem[Hall, Clarke, \& Pringle(1996)]{hal96} Hall, S. M., Clarke, C. J., \& Pringle, J. E. 1996, \aap, 323, 943
\bibitem[Hartmann, Ballesteros-Paredes \& Bergin(2001)]{har01} Hartmann, L., Ballesteros-Paredes, J., \& Bergin, E. A. 2001, \apj, 562, 852
\bibitem[Heitsch, Mac Low, \& Klessen(2001)]{hei01} Heitsch, F.,
Mac Low, M.-M., \& Klessen, R. S. 2001, \apj, 547, 280 (Paper II)
\bibitem[Hogerheide(2001)]{hog01} Hogerheide, M.R. 2001, \apj, 553, 618
\bibitem[Jayawardhana et al.(2001)]{jay01} Jayawardhana, R., Wolk, S. J., Barrado y Navascu\'{e}s D., Telesco, C. M., \& Hearty, T. J. 2001, \apjl, 550, L197
\bibitem[Kim, Ostriker, \& Stone (2003)]{kim03} Kim, W-T, Ostriker, E. C., \& Stone, J. M. 2003, \apj. 595, 574
\bibitem[Klessen(2001)]{kle01} Klessen, R. S. 2001, \apj, 556, 837
\bibitem[Klessen \& Burkert(2000)]{kle00a} Klessen, R. S., \& Burkert, A. 2000, \apjs, 128, 287
\bibitem[Klessen, Heitsch, \& Mac Low(2000)]{kle00b} Klessen, R. S., Heitsch, F., \& Mac Low, M.-M. 2000, \apj, 535, 887 (Paper I)
\bibitem[Koerner, Sargent, \& Beckwith(1993)]{koe93} Koerner, D.W., Sargent, A. I., \& Beckwith, S. V. W. 1993, Icarus, 106, 2
\bibitem[Kolmogorov(1941)]{kol41} Kolmogorov, A. N. 1941, Dokl. Akad. Nauk SSSR, 30, 301
\bibitem[Kritsuk \& Norman(2002)]{kri02} Kritsuk, A. G. \& Norman, M. L. 2002, \apj, 580, L51
\bibitem[Kroupa(2002)]{kro02} Kroupa, P. 2002, Science, 295, 82
\bibitem[Kroupa \& Burkert(2001)]{kro01} Kroupa, P. \& Burkert, A. 2001, \apj, 555, 945
\bibitem[Lazarian, Pogosyan, \& Esquivel(2002)]{laz02} Lazarian, A., Pogosyan, D., \& Esquivel, A. 2002, in ASP Conf. Ser. 276, See Through the Dust, eds. By A. R. Taylor, T. L. Landecker, \& A. G. Willis (San Francisco: ASP), 182
\bibitem[Lejeune \& Bastien(1986)]{lej86} Lejeune, C. \& Bastien, P. 1986, \apj, 309, 167
\bibitem[Lynden-Bell \& Pringle(1974)]{lyn74} Lynden-Bell, D. \& Pringle, J. E. 1974, \mnras, 168, 603
\bibitem[Mac Low(1999)]{mac99} Mac Low, M.-M. 1999, \apj, 524, 169
\bibitem[Mac Low \& Klessen(2004)]{mac04} Mac Low, M.-M. \& Klessen, R. S. 2004, Rev. Mod. Phys., in press
\bibitem[Mac Low et al.(1998)]{mac98} Mac Low, M.-M., Klessen, R. S., Burkert, A., \& Smith, M. D. 1998, \prl, 80, 2754
\bibitem[McKee(1989)]{mck89} McKee, C. F. 1989, \apj, 345, 782
\bibitem[McKee(1999)]{mck99} McKee, C. F. 1999, in The Origin of Stars and Planetary Systems, eds. C. J. Lada \& N. D. Kylafis (Dordrecht: Kluwer), 29
\bibitem[McKee \& Zweibel(1995)]{mck95} McKee, C. F. \& Zweibel, E. G. 1995, \apj, 440, 686
\bibitem[Mestel \& Spitzer(1956)]{mes56} Mestel, L. \& Spitzer, L., Jr., 1956, \mnras, 116, 503
\bibitem[Motte, Andr\'{e}, \& Neri(1998)]{mot98} Motte, F., Andr\'{e}, P., \& Neri, R. 1998, \aap, 336, 150
\bibitem[Mouschovias(1991)]{mou91} Mouschovias, T. Ch. 1991, Single-stage fragmentation and a modern theory of star formation.  In The Physics of Star Formation and Early Stellar Evolution, eds. C. J. Lada \& N. D. Kylafis (Dordrecht:  Kluwer), 449.
\bibitem[Mouschovias \& Paleologou(1979)]{mou79} Mouschovias, T. Ch. \& Paleologou, E. V. 1979, \apj, 230, 204
\bibitem[Mouschovias \& Paleologou(1980)]{mou80} Mouschovias, T. Ch. \& Paleologou, E. V. 1979, \apj, 237, 877
\bibitem[Mouschovias \& Spitzer(1976)]{mou76} Mouschovias, T. Ch., \& Spitzer, L., Jr. 1976, \apj, 210, 326
\bibitem[M\"{u}ller \& Biskamp(2000)] {mul00} M\"{u}ller, W.-C., Biskamp, D. 2000, \prl, 84, 475
\bibitem[Murray \& Lin(1996)]{mur96} Murray, S. D. \& Lin, D. N. C. 1996, \apj, 467, 728
\bibitem[Myers, Evans, \& Ohashi(2000)]{mye00} Myers, P. C., Evans, N. J., \& Ohashi, N. 2000, in Protostars and Planets IV, eds. V. Mannings, A. P. Boss, \& S. S. Russell (Tucson: Univ. Arizona Press), 217
\bibitem[Nakano(1998)]{nak98} Nakano, T. 1998, \apj, 494, 587
\bibitem[Nordlund \& Padoan(1999)]{nor99} Nordlund, ${\rm \AA}$. \& Padoan, P. 1999, in Interstellar Turbulence, eds. J. Franco \& A. Carramiñana (Cambridge: Cambridge Univ. Press), 218
\bibitem[Norman(2000)]{nor00} Norman, M. L. 2000, RevMexAA Conf. Ser., 9, 66
\bibitem[Norman \& Ferrara(1996)]{nor96} Norman, C. A. \& Ferrara, A. 1996, \apj. 467, 280
\bibitem[Ostriker(1998)]{ost98} Ostriker, E. C. 1998, Theory of Protostellar Accretion.  In Accretion Processes in Astrophysical Systems, eds. S. Holt \& T. Kallman (New York: AIP), 484
\bibitem[Ostriker, Gammie, \& Stone(1999)]{ost99} Ostriker, E. C., Gammie, C. F., \& Stone, J. M. 1999, \apj, 513, 259
\bibitem[Padoan \& Nordlund(1999a)]{pad99} Padoan, P. \& Nordlund, ${\rm \AA}$. 1999, \apj, 526, 279
\bibitem[Padoan \& Nordlund(1999b)]{pan99} Padoan, P. \& Nordlund, ${\rm \AA}$. 1999, in Interstellar Turbulence, eds. J. Franco \& A. Carramiñana, (Cambridge: Cambridge Univ. Press), 248
\bibitem[Padoan \& Nordlund(2002)]{pad02} Padoan, P. \& Nordlund, ${\rm \AA}$. 2002, \apj, 576, 870
\bibitem[Padoan, Nordlund, \& Jones(1997)]{pad97} Padoan, P., Nordlund, ${\rm \AA}$., \& Jones, B. J. 1997, \mnras, 288, 145
\bibitem[Padoan et al.(2001)]{pad01} Padoan, P., Nordlund, ${\rm \AA}$., R\"{o}gnvaldsson, \"{O}. E., \& Goodman, A. A. 2001, in ASP Conf. Ser. 243, From Darkness to Light: Origin and Evolution of Young Stellar Clusters, eds. T. Montmerle, T. \& P. Andr\'{e} (San Francisco: ASP), 279
\bibitem[Richer \& Padman(1991)]{ric91} Richer, J. S. \& Padman, R. 1991, \mnras, 251, 707
\bibitem[Salpeter(1955)]{sal55} Salpeter, E. E. 1955, \apj, 121, 161
\bibitem[Sargent \& Beckwith(1991)]{sar91} Sargent, A. I. \& Beckwith, S. V. W. 1991, \apjl, 382, L31
\bibitem[Scally \& Clarke(2001)]{sca01} Scally, A. \& Clarke, C. 2001, \mnras, 325, 449
\bibitem[Scalo et al.(1998)]{sca98} Scalo, J. M., Vázquez-Semadeni, E., Chappell, D., \& Passot, T. 1998, \apj, 504, 835
\bibitem[Sellwood \& Balbus (1999)]{sel99} Sellwood, J. A. \& Balbus, S. A. 1999, \apj, 511, 660
\bibitem[Shu(1977)]{shu77} Shu, F. H. 1977, \apj, 214, 488
\bibitem[Shu, Adams, \& Lizano(1987)]{shu87} Shu, F. H., Adams, F. C., \& Lizano, S. 1987, \araa, 25, 23.
\bibitem[Silk \& Takahashi(1979)]{sil79} Silk, J., \& Takahashi, T. 1979, \apj, 229, 242
\bibitem[Testi \& Sargent(1998)]{tes98} Testi, L. \& Sargent, A. I. 1998, \apjl, 508, L91
\bibitem[Truelove et al.(1997)]{tru97} Truelove, J. K. , Klein, R. I., McKee, C. F., Holliman, J. H. II, Howell, L. H., \& Greenough, J. 1997, \apjl, 489, L179
\bibitem[V\'{a}zquez-Semadeni(1994)]{vaz94} Vázquez-Semadeni, E. 1994, \apj, 423, 681
\bibitem[V\'{a}zquez-Semadeni, Passot, \& Pouquet(1995)]{vaz95} V\'{a}zquez-Semadeni, E., Passot, T., \& Pouquet, A. 1995, \apj, 441, 702 
\bibitem[V\'{a}zquez-Semadeni, Passot, \& Pouquet(1996)]{vaz96} V\'{a}zquez-Semadeni, E., Passot, T., \& Pouquet, A. 1996, \apj, 473, 881 
\bibitem[V\'{a}zquez-Semadeni et al.(2000)]{vaz00} V\'{a}zquez-Semadeni, E., et al. 2000, in Protostars and Planets IV, eds. V. Mannings, A. Boss \& S. Russell (Tucson: Univ. of Arizona Press), 3
\bibitem[Vestuto, Ostriker, \& Stone(2003)]{ves03} V\'{a}zquez-Semadeni, E., Ostriker, E. C., Passot, T., Gammie, C. F., \& Stone, J.M. 2000, in Protostars and Planets IV, eds. V. Mannings, A. Boss \& S. Russell (Tucson: Univ. of Arizona Press), 3
\bibitem[Williams, Blitz, \& McKee(2000)]{wil00} Williams, J. P., Blitz, L., \& McKee, C.F. 2000, in Protostars and Planets IV, eds. V. Mannings, A. Boss \& S. Russell (Tucson: Univ. of Arizona Press), 97
\bibitem[Williams, De Geus, \& Blitz(1994)]{wil94} Williams, J. P., De Geus, E. J.,  \& Blitz, L. 1994, \apj, 428, 693
\bibitem[Williams et al.(1999)]{wil99} Williams, J. P., Myers, P.C., Wilner, D. J., Di Francesco, J., \& Tafalla, M. 1999, \apjl, 513, L61
\bibitem[Wiseman et al.(2001)]{wis01} Wiseman, J. J., Wootten, A., Zinnecker, H., \& McCaughrean, Mark. 2001, \apjl, 550, L87
\bibitem[Zhang, Hunter, \& Sridharan(1998)]{zha98} Zhang, Q., Hunter, T. R., \& Sridharan, T. K. 1998, \apjl, 505, L151
\end{thebibliography}
\end{document}